\documentclass{article}

\usepackage{arxiv}

\usepackage[utf8]{inputenc} 
\usepackage[T1]{fontenc}    
\usepackage{hyperref}       
\usepackage{url}            
\usepackage{booktabs}       
\usepackage{amsfonts}       
\usepackage{nicefrac}       
\usepackage{microtype}      
\usepackage{lipsum}
\usepackage{graphicx}
\graphicspath{ {./images/} }

\usepackage{tikz,pgfplots}
\usepackage{graphicx}
\usepackage{graphics}
\usepackage{float}
\usepackage{booktabs}
\usepackage{multirow}
\usepackage{amsmath}
\usepackage{subfig}

\title{A highly scalable numerical framework for reservoir simulation on UG4 platform}

\author{
 Shuai Lu \\
  Computer, Electrical and Mathematical Sciences and Engineering Division,\\
  King Abdullah University of Science and Technology,\\
  Thuwal, 23955-6900, Saudi Arabia \\
  \texttt{shuai.lu@kaust.edu.sa} \\
}

\begin{document}
\maketitle
\begin{abstract}
The modeling and simulation of multiphase fluid flow receive significant attention in reservoir engineering. Many time discretization schemes for multiphase flow equations are either explicit or semi-implicit, relying on the decoupling between the saturation equation and the pressure equation. In this study, we delve into a fully coupled and fully implicit framework for simulating multiphase flow in heterogeneous porous media, considering gravity and capillary effects. We utilize the Vertex-Centered Finite Volume Method for spatial discretization and propose an efficient implementation of interface conditions for heterogeneous porous media within the current scheme. Notably, we introduce the Linearly Implicit Extrapolation Method (LIMEX) with an error estimator, adapted for the first time to multiphase flow problems. To solve the resulting linear system, we employ the BiCGSTAB method with the Geometric Multigrid (GMG) preconditioner. The implementations of models and methods are based on the open-source software: UG4. The results from parallel computations on the supercomputer demonstrate that the scalability of our proposed framework is sufficient, supporting a scale of thousands of processors with Degrees of Freedom (DoF) extending up to billions.
\end{abstract}


\section{Introduction}
In order to understand the flow of fluids in geological layers, lots of reservoir simulation models are developed \cite{kazemi1978efficient, coats1980equation, helmig1997multiphase, cao2002development,  moortgat2011compositional, shao2021predicting}.
These advection, diffusion, or even reaction processes are usually described as
multi-phase flows in porous media \cite{bastian1999parallel}. Most of the models are diffuse interface multiphase flow models where the interfaces between phases are captured by numerical schemes as diffuse zones, and the apparition or disappearance of interfaces are naturally obtained \cite{sun2020reservoir}.\\
After modeling, it is significant to apply the proper spatial discretization scheme to approximate the physical fields. Finite Difference Methods(FDM) are popular and efficient for simulations on regular domains \cite{peaceman2000fundamentals}. Compared to FDM, Finite Volume Methods(FVM) offer more flexibility on unstructured grids. Furthermore, FVM is competitive due to its mass conservation and monotonicity properties \cite{michel2003finite, reichenberger2006mixed}. Mixed and mixed-hybrid finite element methods are also popular for advection-dominated flow problems\cite{huber2000node}. However, they may not be so favorable for capillary pressure-dominated flow problems\cite{huber1999multiphase}.\\
After spatial discretization, employing millions or even billions of degrees of freedom is a common practice in field-scale flow and transport simulations to represent complex geological heterogeneity. Therefore, the use of parallel reservoir simulators with robust, efficient, and scalable algorithms is crucial for addressing the challenges associated with large-scale geological models in the field of reservoir simulation. The numerical simulator for incompressible two-phase flow in porous media was developed by Douglas et al \cite{douglas1959method} in 1959. Numerous simulators and schemes were developed after that. One of the most famous methods, Implicit Pressure, Explicit Saturation (IMPES) scheme, was developed by Sheldon et al. \cite{sheldon1959one} and Stone et al. \cite{stone1961analysis} and has been widely used in multi-phase flow solvers. The base assumption of IMPES is that the pressure depends on the saturation weakly. Therefore, the calculation of pressure is separated from that of saturation. The computation cost on each time step is cheap because the system of governing equations is split and solved sequentially. However, the explicit scheme for solving the saturation equations always leads to severe numerical stability \cite{coats2003impes}. Therefore, it requires quite small time step sizes to satisfy Courant–Friedrichs–Lewy (CFL) condition, which makes it not an alternative choice for the reservoir simulation over a long period of time. Compared to the semi-implicit schemes, the fully implicit scheme is more reliable. In such
schemes, the time derivatives in the governing equations are discretized implicitly and all the spatial derivatives are evaluated at the new time step. The fully implicit scheme is more stable and can be used for the simulation with larger time step sizes. Apart from choosing different time stepping schemes, the decoupling of highly nonlinear equations also results in a severe time step restriction. The strength of the coupling between pressure and saturation equation depends on the applied formulation, it is weaker for the global pressure formulation \cite{bastian1999numerical}. While the spatial variability of rock properties(permeability and porosity) and constitutive relations(relative permeability, capillary pressure) strengthen the coupling. Therefore, a fully coupled and fully implicit discretization of the governing equations is considered the best choice to maximize robustness\cite{bastian1999numerical, dawson1997parallel, luo2020fully}.\\
There is no doubt that Newton-like methods are most popular for solving nonlinear equations. Dawson et al. used the mixed finite element method for spatial discretization combined with Implicit Euler for temporal discretization \cite{dawson1997parallel}, where Newton-Krylov method is employed for nonlinear iterations and the Generalized Minimal Residual method(GMRES) is used for transferring the Krylov information. Recently, a scalable sequential fully implicit framework is proposed by Yang et al.\cite{yang2018scalable}. The great convergences are achieved by using Newton-Krylov method with an additive Schwarz preconditioner. However, highly nonlinear problems always take too many Newton-like iterations to get a converged solution. In order to get rid of nonlinear iterations, linearly implicit methods come to the stage. They can be classified as two main types. One type is Rosenbrock-Wanner methods(ROW), where an exact Jacobian matrix is needed. Another type is W-methods, where only an approximation of Jacobian matrix is required. The linearly implicit extrapolation method(LIMEX)\cite{deuflhard1983order, deuflhard1985recent, deuflhard1987extrapolation} can be classified as W-method, but eliminates the requirement of algebraic conditions for coefficients. It requires two loops: the outer loop for adaptive time step discretization and the inner loop for the numerical solution of the arising linear systems\cite{deuflhard2012adaptive}. Recently, the efficiency of applying LIMEX in the numerical solutions of density driven flow problems has been investigated\cite{nagel2018efficient}.\\
In this paper, a highly scalable framework is proposed for multiphase flow problem in porous media. The Vertex Centered Finite Volume Method with upwind scheme is employed for the spatial discretization. The efficient implementation of interface conditions of heterogeneous porous media is proposed accordingly. The LIMEX scheme with the error estimator is adapted for two-phase flow problems. The arising linear system is solved by BiCGSTAB with the Geometric Multigrid(GMG) preconditioner. The implementations of models and methods are based on the open-source software: UG4.\\
The organization of this work is as follows. The mathematical model for multiphase flow is introduced in Section \ref{modeling}. The spatial discretization schemes, including the upwind scheme and the interface condition, are illustrated in Section \ref{Spatial}. In Section \ref{stepping}, the linearly implicit extrapolation scheme with the error estimator is adapted for the fully coupled framework. The validation and scalability of the proposed framework are present in Section \ref{tests}. Finally, the conclusions are drawn.\\

\section{Mathematical Model}\label{modeling}
For a domain $\Omega \subset \mathbb{R}^d$ and time interval $\mathcal{T}=(0, T)$, the general mass conservation equations of each phase $\alpha$ in the porous medium is given by
\begin{equation} \label{m}
	\frac{\partial\left(\Phi \rho_\alpha S_\alpha\right)}{\partial t}+\nabla \cdot\left\{\rho_\alpha \mathbf{u}_\alpha\right\}=q_\alpha
\end{equation}
where $\Phi$ is the porosity of the porous medium, $\rho_\alpha$, $S_\alpha$ and $q_\alpha$ are the density, saturation and source of phase $\alpha$, respectively.
The velocity $\mathbf{u}_\alpha$ can be obtained from the extended Darcy's Law as
\begin{equation} \label{u}
	\mathbf{u}_\alpha=-\frac{k_{r \alpha}}{\mu_\alpha} \mathbf{K}\left(\nabla p_\alpha-\rho_\alpha \mathbf{g}\right)
\end{equation}
where $k_{r \alpha}$, $\mu_\alpha$ and $p_\alpha$ are relative permeability, viscosity and pressure, respectively. $\mathbf{K}$ is the absolute permeability tensor and $\mathbf{g}$ is gravity vector. All the accessible space of the porous medium is assumed to be saturated by fluids, so the sum of all the saturations is 1:
\begin{equation} \label{S}
	\sum_\alpha S_{\alpha}=1,\quad 0\leq S_\alpha\leq1
\end{equation}
The relation between different phase pressure is described by the capillary pressure:
\begin{equation} \label{pc}
	p_{c \alpha\beta }=p_\beta-p_\alpha \quad \forall \beta \neq \alpha
\end{equation}
In this paper, we discuss the in-compressible immiscible two-phase flow in details. There are one wetting phase $w$ and one no-wetting phase $n$. The pressure of no-wetting phase $p_n$ and the saturation of wetting phase $S_w$ are chosen as the primary variables. Insert Equations (\ref{u}, \ref{S}, \ref{pc}) into Equations (\ref{m}), one get two mass conservation equations:
\begin{equation} \label{sW}
\Phi\rho_w \frac{\partial S_w}{\partial t}+\rho_w\nabla \cdot\left\{ \frac{ k_{r w} p_c^{'}}{\mu_w}\mathbf{K} \nabla S_w +k_{r w}\mathbf{v}_w \right\}=q_{w}
\end{equation}
\begin{equation} \label{pN}
\Phi \rho_n\frac{\partial \left(1-S_w\right)}{\partial t}+\rho_n\nabla \cdot\left\{ k_{rn} \mathbf{v}_n\right\}=q_{n}
\end{equation}
where $p_c$, $k_{r w}$ and $k_{r n}$ are the functions of $S_w$. $p_c^{'}$ is the derivative of $p_c$ wrt. $S_w$. Here 
\begin{equation} \label{vW}
	\mathbf{v}_w=-\frac{\mathbf{K}}{\mu_w}(\nabla p_n-\rho_w \mathbf{g})
\end{equation}
\begin{equation} \label{vN}
	\mathbf{v}_n=-\frac{\mathbf{K}}{\mu_n}(\nabla p_n-\rho_n \mathbf{g})
\end{equation}
Please note that $\mathbf{v}_w$ is not a directional vector of $\mathbf{u}_w$ since the capillary diffusion term is separated by substituting $p_w$ into $\mathbf{u}_w$. In in-compressible case, $\Phi$ and $\rho_n$ are constant. The Dirichlet and Neumann boundary conditions for primary variables are given by
\begin{equation} \label{BsWd}
	S_w(\mathbf{x}_d,t) = S_{wd}(\mathbf{x}_d) \quad \forall \mathbf{x}_d \in \partial\Omega_{S_{wd}}
\end{equation}
\begin{equation} \label{BsWn}
	\left\{ \frac{\rho_w k_{r w} p_c^{'}}{\mu_w}\mathbf{K} \nabla S_w +\rho_w k_{r w}\mathbf{v}_w \right\} \cdot \mathbf{n} = \phi_{s}(\mathbf{x}_n) \quad \forall \mathbf{x}_n \in \partial\Omega_{S_{wn}}
\end{equation}
\begin{equation} \label{BpNd}
	p_n(\mathbf{x}_d,t) = p_{nd}(\mathbf{x}_d) \quad \forall \mathbf{x}_d \in \partial\Omega_{p_{nd}}
\end{equation}
\begin{equation} \label{BpNn}
	k_{r n}\mathbf{v}_n \cdot \mathbf{n} = \phi_{p}(\mathbf{x}_n) \quad \forall \mathbf{x}_n \in \partial\Omega_{p_{nn}}
\end{equation}
\section{Spatial discretization}\label{Spatial}
\subsection{Vertex Centered Finite Volume Method}
\begin{figure}[H]
	\begin{center}
		\begin{tikzpicture}
			\draw[thick] (0,0) -- (0,2) -- (2,0)-- (0,0);
			\draw[thick] (0,0) -- (0,2) -- (-2.5,0)--(0,0);
			\draw[thick] (0,0) -- (-2.5,0) -- (-2.3,-1.6) --(0,-1.8) --(0,0);
			\draw[thick] (0,0) -- (0,-1.8) -- (2,0) --(0,0);
			\draw[fill] (0,0) circle(0.6mm) node[above right]{$v_i$};
			\draw[fill] (-2.5,0) circle(0.6mm) node[above left]{$v_j$};
			\draw[fill] (2,0) circle(0.6mm); 
			\draw[fill] (0,2) circle(0.6mm); 
			\draw[fill] (0,-1.8) circle(0.6mm); 
			\draw[fill] (-2.3,-1.6) circle(0.6mm); 
			\draw (-1.12,-0.82) circle(0.6mm); 
			\draw (-0.8,0.6) circle(0.6mm); 
			\draw (0.67,0.6) circle(0.6mm); 
			\draw (0.67,-0.6) circle(0.6mm); 
			
			\draw[dashed] (-1.12,-0.82) -- (-1.25,0);
			\draw[dashed] (-1.12,-0.82) -- (0,-0.9);
			\draw[dashed] (-1.12,-0.82) -- (-1.15,-1.7);
			\draw[dashed] (-1.12,-0.82) -- (-2.4,-0.8);
			
			\draw[dashed] (-0.8,0.6) -- (-1.25,0);
			\draw[->,semithick]	(-1.025,0.3) -- (-1.5,0.68);
			\draw (-1.025,0.3) ++ (0.1, 0.12) -- ++ (-0.119, 0.09) -- ++ (-0.1, -0.133);
			\draw (-1.05,0.55) node[above left]{$\mathbf{n}$};
			\draw[->,semithick]	(-1.025,0.3) -- (-1.6,0.5);
			\draw (-1.25,0.5) node[below left]{$\mathbf{v}_w$};
			
			\draw[dashed] (-0.8,0.6) -- (0,0.9);
			\draw[dashed] (-0.8,0.6) -- (-1.25,1);
			
			\draw[dashed] (0.67,0.6) -- (1,0);
			\draw[dashed] (0.67,0.6) -- (0,0.9);
			\draw[dashed] (0.67,0.6) -- (1,1);
			
			\draw[dashed] (0.67,-0.6) -- (1,0);
			\draw[dashed] (0.67,-0.6) -- (0,-0.9);
			\draw[dashed] (0.67,-0.6) -- (1,-0.9);
		\end{tikzpicture}
		\caption{2D mesh for vertex-centered finite volume method}
		\label{fvm}
	\end{center}
\end{figure}
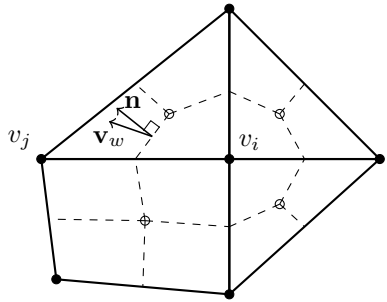

Equations (\ref{sW}, \ref{pN}) are discretized on an unstructured mesh $E_h=\left\{e_1, \ldots, e_{N_e}\right\}$, which covers the domain $\Omega$. The set of vertices is $V=\left\{v_1, \ldots, v_{N_v}\right\}$. The control volume of each vertices is constructed as the secondary mesh $B_h=\left\{b_1, \ldots, b_{N_v}\right\}$. The example of 2D mesh are shown in Figure \ref{fvm}, where the finite elements are constructed by solid lines.
The control volume of $v_i$ is consisted of its surrounding dashed lines. The discretized forms of Equations (\ref{sW}, \ref{pN}) at $v_i$ are:
\begin{eqnarray}\label{dsW}
        \nonumber
		\frac{\partial\left(\sum_{N \in e_j\cap b_i} \left.\Phi_j \rho_{w}S_{w }\right|_{\mathbf{x}_N} A_{b^j_i}\right)}{\partial t}
		+
		\sum_{M \in e_j\cap \partial b_i} \left.\frac{\rho_w k_{r w} p_c^{'}}{\mu_w}\mathbf{K} \nabla S_w\mathbf{n} \right|_{\mathbf{x}_M}l_M		
		+\sum_{M \in e_j\cap \partial b_i} \left.\rho_w k_{r w}\mathbf{v}_w \mathbf{n} \right|_{\mathbf{x}_M}l_M \\
		=\sum_{N \in e_j\cap b_i} \left.q_{w}\right|_{\mathbf{x}_N} A_{b^j_i}
\end{eqnarray}
\begin{equation}\label{dpN}
	\frac{\partial\left(\sum_{N \in e_j\cap b_i} \left.\Phi_j \rho_{n}\left(1-S_w\right)\right|_{\mathbf{x}_N} A_{b^j_i}\right)}{\partial t}	
	+\sum_{M \in e_j\cap \partial b_i} \left.\rho_n k_{rn}\mathbf{v}_n \mathbf{n} \right|_{\mathbf{x}_M}l_M
	=\sum_{N \in e_j\cap b_i} \left.q_{n}\right|_{\mathbf{x}_N} A_{b^j_i}
\end{equation}

where $\mathbf{x}_N$ is the barycenter of sub-control volume $b^j_i$, $A_{b^j_i}$ is the area of sub-control volume face, $\mathbf{x}_M$ is the midpoint of sub-control volume edge, $l_M$ is the length of sub-control volume edge.
\subsection{Upwind scheme}
Multi-phase problem could be either diffusion or convection-dominated, which depends on parameters. Therefore, a upwind scheme is expected to eliminate numerical oscillation, especially for convection-dominated cases. Different from the upwind scheme where Darcy velocities $\mathbf{u}_\alpha$ are utilized in the upwind criteria, the velocity  $\mathbf{v}_\alpha$ are used in the present upwind scheme. This modification prevents the capillary diffusion part $\frac{ k_{r w} p_c^{'}}{\mu_w}\mathbf{K} \nabla S_w$ from being upwind. The upwind evaluations for the coupled Equations (\ref{dsW}, \ref{dpN}) are as follows:
\begin{eqnarray}\label{dsW}
    \nonumber
	\rho_{w}\frac{\partial\left(\sum_{N \in e_j\cap b_i} \left.\Phi_j S_{w }\right|_{\mathbf{x}_N} A_{b^j_i}\right)}{\partial t}
	+
	\rho_w\sum_{M \in e_j\cap \partial b_i} \left.\frac{ k_{r w} p_c^{'}}{\mu_w}\mathbf{K} \nabla S_w\mathbf{n} \right|_{\mathbf{x}_M}l_M		
	+
	\rho_w\sum_{M \in e_j\cap \partial b_i} \left. \left. k_{rw}\right|_{\mathbf{x}_{up}}\mathbf{v}_w \mathbf{n} \right|_{\mathbf{x}_M}l_M\\
	=\sum_{N \in e_j\cap b_i} \left.q_{w}\right|_{\mathbf{x}_N} A_{b^j_i}
\end{eqnarray}
\begin{equation}\label{dpN}
	\rho_{n}\frac{\partial\left(\sum_{N \in e_j\cap b_i} \left.\Phi_j \left(1-S_w\right)\right|_{\mathbf{x}_N} A_{b^j_i}\right)}{\partial t}	
	+\rho_{n}\sum_{M \in e_j\cap \partial b_i} \left. \left.k_{r n}\right|_{\mathbf{x}_{up}}\mathbf{v}_n \mathbf{n} \right|_{\mathbf{x}_M}l_M
	=\sum_{N \in e_j\cap b_i} \left.q_{n}\right|_{\mathbf{x}_N} A_{b^j_i}
\end{equation}
where $\mathbf{x}_{up}$ is the upwind node, $\left.()\right|_{\mathbf{x}_{up}}$ is evaluated as follows
\begin{equation}\label{up_krw}
	\left. k_{rw}\right|_{\mathbf{x}_{up}}=
	\begin{cases}
		k_{rw}(\left.S_w\right|_{\mathbf{x}_i}) & \left.\mathbf{v}_w \cdot \mathbf{n}\right|_{\mathbf{x}_M} \geq 0 \\
		k_{rw}(\left.S_w\right|_{\mathbf{x}_j}) & \text { else }
	\end{cases}
\end{equation}
\begin{equation}\label{up_krn}
\left.k_{r n}\right|_{\mathbf{x}_{up}}=
\begin{cases}
	k_{rn}(\left.S_w\right|_{\mathbf{x}_i}) & \left.\mathbf{v}_n \cdot \mathbf{n}\right|_{\mathbf{x}_M} \geq 0 \\
	k_{rn}(\left.S_w\right|_{\mathbf{x}_j}) & \text { else }
\end{cases}
\end{equation}

\subsection{heterogeneity}
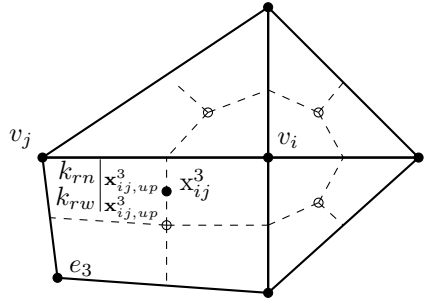
\begin{figure}[H]
	\begin{center}
		\begin{tikzpicture}
			\draw[thick] (0,0) -- (0,2) -- (2,0)-- (0,0);
			\draw[thick] (0,0) -- (0,2) -- (-3,0)--(0,0);
			\draw[thick] (0,0) -- (-3,0) -- (-2.8,-1.6) --(0,-1.8) --(0,0);
			\draw[thick] (0,0) -- (0,-1.8) -- (2,0) --(0,0);
			\draw[fill] (0,0) circle(0.6mm) node[above right]{$v_i$};
			\draw[fill] (-3,0) circle(0.6mm) node[above left]{$v_j$};
			\draw[fill] (2,0) circle(0.6mm); 
			\draw[fill] (0,2) circle(0.6mm); 
			\draw[fill] (0,-1.8) circle(0.6mm); 
			\draw[fill] (-2.8,-1.6) circle(0.6mm); 
			\draw (-1.35,-0.9) circle(0.6mm); 
			\draw (-0.8,0.6) circle(0.6mm); 
			\draw (0.67,0.6) circle(0.6mm); 
			\draw (0.67,-0.6) circle(0.6mm); 
			
			\draw[dashed] (-1.35,-0.9) -- (-1.35,0);
			\draw[dashed] (-1.35,-0.9) -- (0,-0.9);
			\draw[dashed] (-1.35,-0.9) -- (-1.35,-1.7);
			\draw[dashed] (-1.35,-0.9) -- (-2.9,-0.8);
			
			\draw[dashed] (-0.8,0.6) -- (-1.35,0);
			
			\draw[dashed] (-0.8,0.6) -- (0,0.9);
			\draw[dashed] (-0.8,0.6) -- (-1.25,1);
			
			\draw[dashed] (0.67,0.6) -- (1,0);
			\draw[dashed] (0.67,0.6) -- (0,0.9);
			\draw[dashed] (0.67,0.6) -- (1,1);
			
			\draw[dashed] (0.67,-0.6) -- (1,0);
			\draw[dashed] (0.67,-0.6) -- (0,-0.9);
			\draw[dashed] (0.67,-0.6) -- (1,-0.9);

            \draw (-1.3,-0.7) node[above right]{$\mathrm{x}_{ij}^3$};
            \filldraw[draw=black,fill=black] (-1.35,-0.45) circle (0.6mm);
            \draw (-1.3,-0.6) node[above left]{$k_{rn}|_{\mathbf{x}_{ij,up}^3}$};
            \draw (-1.3,-0.95) node[above left]{$k_{rw}|_{\mathbf{x}_{ij,up}^3}$};
            \draw (-2.2,-1.7) node[above left]{$e_3$};
		\end{tikzpicture}
		\caption{The evaluations of the relative permeabilities at the integral point $x_{ij}^3$}
		\label{fvm2}
	\end{center}
\end{figure}

Based on the extended capillary pressure condition \cite{van1995effect}, the general interface conditions at media discontinuities for vertex-centered finite volume method can be developed by the competition of capillary pressure. The evaluation of saturation at vertex $v_i$ wrt. element $e_j$ depends on $p_{\text {cmin }, i}$, which is defined as following
\begin{equation}\label{pcmin}
	p_{\text {cmin }, i}=\min _{k \in E_i} p_c\left(\mathbf{x}^k\right)
\end{equation}
where $E_i$ are the elements having vertex $v_i$, $\mathbf{x}^k$ is the barycenter of element $e_k$. However, this scheme is not efficient for large scale parallel computing, since the computation of $p_{\text {cmin }, i}$ is expensive. The dynamic update of $p_{\text {cmin }, i}$ is required with the changing of saturation. Moreover, for each update, extra communication between processors is needed if elements $E_i$ are not distributed on the same processor. A more efficient scheme is proposed by Bastian \cite{bastian1999numerical} as following
\begin{equation}\label{pcmin2}
	p_{\text {cmin }, i}=\min _{k \in e_i} p_c\left(\mathbf{x}^k, 1-S_{n, i}\right)
\end{equation}
where $p_c$ is not evaluated at $\mathbf{x}^k$ restrictively, since the saturation of vertex $S_{n, i}$ are used for every evaluate point $\mathbf{x}^k$ in elements $e_i$.\\
In the current framework, a scheme without the calculation of $p_c$ is implemented, which depends on $p_{\text {dmin }}$. The definition of $p_{\text {dmin }, i}$ at vertex $v_i$ is as following
\begin{equation}\label{pdmin}
	p_{\text {dmin }, i}=\min _{N \in e_j\cap b_i} p_{\text {d}}(N)
\end{equation}
where $p_{\text {d}}(N)$ is the entry pressure of element $e_j$, $N \in e_j\cap b_i$. The saturation is evaluated by
\begin{equation}\label{hSw}
	\hat{S}_{w,i,N}=
	\begin{cases}
		S_{w,i} & p_{\text {d}}(N)=p_{\text {dmin }, i} \\
		1-S_{n,r} & p_{\text {d}}(N)\geq p_{\text {dmin }, i}J(S_{w,i}) \\
		S & \text{where} \; S \; \text{solves}\; p_{\text {d}}(N)J(S) = p_{\text {dmin }, i}J(S_{w,i})  \\
	\end{cases}
\end{equation}
where $\hat{S}_{w,i,N}$ is the saturation evaluated in the sub control volume $e_j\cap b_i$, $S_{w,i}$ is the saturation at vertex $v_i$, $J$ is the Leverett-J function \cite{leverett1941capillary} for Brooks-Corey capillary pressure function. Notice that Equation (\ref{hSw}) should be applied in the upwind scheme as well. However, such barrier conditions are only for the non-wetting phase and are not imposed on the convection terms of the wetting phase equations (\ref{dsW}). For instance, the evaluations of the relative permeabilities at $x_{ij}^3$ are illustrated in Figure \ref{fvm2}, where 
\begin{equation} \label{krn_ip}
    k_{rn}|_{\mathbf{x}_{ij,up}^3}= k_{rn}(\hat{S}_{w,up}^3)
\end{equation}
and 
\begin{equation} \label{krw_ip}
    k_{rw}|_{\mathbf{x}_{ij,up}^3} = k_{rw}(S_{w,up})
\end{equation}

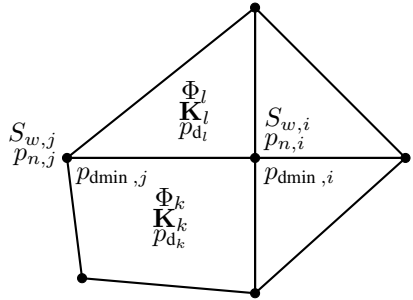
\begin{figure}[H]
	\begin{center}
		\begin{tikzpicture}
			\draw[thick] (0,0) -- (0,2) -- (2,0)-- (0,0);
\draw[thick] (0,0) -- (0,2) -- (-2.5,0)--(0,0);
\draw[thick] (0,0) -- (-2.5,0) -- (-2.3,-1.6) --(0,-1.8) --(0,0);
\draw[thick] (0,0) -- (0,-1.8) -- (2,0) --(0,0);
\draw[fill] (0,0) circle(0.6mm) node[below right]{$p_{\text {dmin }, i}$};
\draw (0,0.2) node[above right]{$S_{w,i}$} node[right]{$p_{n,i}$};
\draw[fill] (-2.5,0) circle(0.6mm) node[above left]{$S_{w,j}$} node[ left]{$p_{n,j}$} node[below right]{$p_{\text {dmin }, j}$};
\draw[fill] (2,0) circle(0.6mm); 
\draw[fill] (0,2) circle(0.6mm); 
\draw[fill] (0,-1.8) circle(0.6mm); 
\draw[fill] (-2.3,-1.6) circle(0.6mm); 
\draw (-1.12,-0.81) node[above]{$\Phi_k$}; 
\draw (-1.12,-0.82) node[]{$\mathbf{K}_k$};
\draw (-1.12,-0.83) node[below]{$p_{\text {d}_k}$};

\draw (-0.8,0.61) node[above]{$\Phi_l$}; 
\draw (-0.8,0.6) node[]{$\mathbf{K}_l$};
\draw (-0.8,0.59) node[below]{$p_{\text {d}_l}$};
		\end{tikzpicture}
		\caption{Variables and constants over the grid}
		\label{variables}
	\end{center}
\end{figure}

In this scheme, $p_{\text {dmin }, i}$ is constant wrt. vertex $v_i$. The variables and constants are shown in Figure \ref{variables}. Where $S_{w}$ and $p_{n}$ are variables. $\Phi$, $\mathbf{K}$ and $p_{\text {d}}$ are constants wrt. each element. While $p_{\text {dmin }}$ is vertex-wise constant and it can be calculated in the pre-processing stage. 
In this scheme, extra communication between processors can be avoided and the scalability of the current framework will not be affected in heterogeneity cases. One can prove that, for $S_w\in C^0$, Equation (\ref{pdmin}) is equivalent to Equation (\ref{pcmin}) when grid size $h$ approaches 0.
\section{Adaptive time stepping scheme}\label{stepping}
The linearly implicit extrapolation method is employed for adaptive temporal discretization in the current framework. After spatial discretization one obtain 
\begin{equation}
	M(u) u'=F(u)
\end{equation}
In linearly implicit extrapolation method, the following temporal discretization is applied
\begin{equation}
	(M(u_{\text {old}})- \tau_{\text {old}}J(u_{\text {old}}))(u_{\text {new}} - u_{\text {old}})=\tau_{\text {old}}F(u_{\text {old}})
\end{equation}
where $u_{\text {new}}$ is the solution for the new step, $u_{\text {old}}$ is the solution of current step, $\tau_{\text {old }}$ is the step size of current step, $J = F_u$ is the exact Jacobi matrix. For a proper adaptive step size, the above equation can be solved by Newton's method in one step.
The adaptive step size is selected from
\begin{equation}
	\tau_{\text {new }}=\sqrt[q+1]{\rho \frac{\mathrm{TOL}}{[[\epsilon]]}} \tau_{\text {old }}
\end{equation}
where $\tau_{\text {new }}$ is the suggested step size for the new step, $q$ is the order of LIMEX scheme, $\rho$ is a safety factor, $\mathrm{TOL}$ is error tolerance, and $[[\epsilon]]$ is the estimate for the relative error of the scaled norm of the solution.
According to the theoretical analysis by Lubich\cite{lubich1995linearly}, the effective order of W-method will be reduced to 2. Therefore the future order of LIMEX is suggested to be fixed to $q=2$\cite{nagel2018efficient}.\\
In order to get the value of $[[\epsilon]]$, the error estimator of the solution $\mathbf{u}(t)=(S_w, p_n)^T(t)$ is needed. Considering the convergence of convection parts of the mass conservation equations, one can get
\begin{equation} \label{norm_uw}
	\begin{aligned}
		&\left\| \rho_w k_{rw}(S_{w1})\mathbf{v}_{w1} -\rho_w k_{rw}(S_{w2})\mathbf{v}_{w2} \right\|\\
		&= \left\| \rho_w\frac{k_{rw}(S_{w1})\mathbf{K}}{\mu_w} \left(\nabla p_{n1}-\rho_w \mathbf{g}\right) - \rho_w\frac{k_{rw}(S_{w2})\mathbf{K}}{\mu_w} \left(\nabla p_{n2}-\rho_w \mathbf{g}\right) \right\|\\
		&\leq  \max \{\left\| \rho_w\frac{k_{rw}(S_w)\mathbf{K}}{\mu_w}\right\|\} \left\|\nabla p_{n1}-\nabla p_{n2}\right\| + \max \{\left\|\rho_w\frac{k_{rw}^{'}(S_w)\mathbf{K}}{\mu_w}(\nabla p_{n}-\rho_w \mathbf{g})\right\|\} \left\|S_{w1}-S_{w2}\right\|
	\end{aligned}
\end{equation}
\begin{equation} \label{norm_un}
	\begin{aligned}
		&\left\| \rho_n k_{rn}(S_{w1})\mathbf{v}_{n1} -\rho_n k_{rn}(S_{w2})\mathbf{v}_{n2} \right\|\\
		&= \left\| \rho_n\frac{k_{rn}(S_{w1})\mathbf{K}}{\mu_n} \left(\nabla p_{n1}-\rho_n \mathbf{g}\right) - \rho_n\frac{k_{rn}(S_{w2})\mathbf{K}}{\mu_n} \left(\nabla p_{n2}-\rho_n \mathbf{g}\right) \right\|\\
		&\leq  \max \{\left\| \rho_n\frac{k_{rn}(S_w)\mathbf{K}}{\mu_n}\right\|\} \left\|\nabla p_{n1}-\nabla p_{n2}\right\| + \max \{\left\|\rho_n\frac{k_{rn}^{'}(S_w)\mathbf{K}}{\mu_n}(\nabla p_{n}-\rho_n \mathbf{g})\right\|\} \left\|S_{w1}-S_{w2}\right\|
	\end{aligned}
\end{equation}
where $\|\mathbf{u}\|:=\sqrt{\int_{\Omega} \mathbf{u}^2}$ is the measure of the energy of velocity field $\mathbf{u}$. From Equation (\ref{norm_uw}) and (\ref{norm_un}), it is obvious that the convergence of $\left\| \nabla p_n(t)\right\|^2$ and $\left\| S_w(t)\right\|^2$ led to the convergence of the convection parts of the mass conservation equations. Based on this fact, the general scaled norm of solution $\mathbf{u}(t)$ can be defined as follows
\begin{equation} \label{norm_u}
		\||\mathbf{u}(t)\||^2 :=\alpha \left\| \nabla p_n(t)\right\|^2 + \beta \left\| S_w(t)\right\|^2
\end{equation}
where $\alpha$ and $\beta$ are the positive coefficients of $\text{H}_1$-seminorm of pressure and $\text{L}_2$-norm of saturation respectively. The error estimator $[[\epsilon]]$ is defined as the relative error of $\||\mathbf{u}(t)\||$. Following Equation (\ref{norm_uw}) and (\ref{norm_un}), $\alpha$ should be much greater than $\beta$, which confirms that multiphase problems are pressure dominated. Conversely, $\beta$ should be large enough to guarantee the accuracy of saturation, as one hopes to keep it monotonic. Therefore, in practice, $\alpha$ and $\beta$ are selected adaptively to balance the error from pressure and saturation.

\section{Numerical results}\label{tests}
The implementation of the schemes is based on the open-source software: UG4 \cite{vogel2013ug,reiter2013massively}. The numerical experiments are carried out on Shaheen III, the supercomputer at King Abdullah University of Science and Technology (KAUST). Shaheen III consisted of 4,608 dual sockets compute nodes based on 96-core AMD Genoa processors running at 2.4GHz. Each node has 384GB of DDR5 memory running at 4800MHz.\\
In Implicit Euler scheme, the absolute(relative) tolerance of the nonlinear iteration is set to $10^{-8}(10^{-6})$. In LIMEX scheme, the error tolerance of the solution $\mathrm{TOL}$ is set to generate comparable solutions. In both time stepping schemes, the linear systems are solved by GMG preconditioned BiCGSTAB method with absolute(relative) tolerance of $10^{-8}(10^{-6})$.\\
To validate the implementation of framework, the numerical test of Buckley-Leverett flow is performed. This flow problem is a kind of Riemann problem, which is very well understood and its exact solution is achieved. More description of Buckley-Leverett can be found in \cite{helmig1997multiphase, bastian1999numerical}. The parameters for the test are provided in Table \ref{case1}.\\
\begin{table}[H]
	\begin{center}
		\caption{Parameters for Case-1}
		\begin{tabular}{*{3}{c}}
			\toprule
			\multicolumn{3}{c}{Case-1: Buckley-Leverett}\\
			\midrule
			{Domain} & \multicolumn{2}{c}{300 m $\times$ 75 m}\\
			{Rock properties} & \multicolumn{2}{c}{$\Phi=0.2$, $K=10^{-7}$ m$^2$}\\
			\multirow{2}*{Fluid properties} & \multicolumn{2}{c}{$\rho_w=\rho_n=1\times10^3$ kg/m$^3$}\\
			& \multicolumn{2}{c}{$\mu_w=\mu_n=1\times10^{-3}$ Pa s}\\
			{Residual saturation} & \multicolumn{2}{c}{$S_{wr}=S_{nr}=0$}\\
			{Capillary pressure} & \multicolumn{2}{c}{$p_c\equiv0$}\\
			{Relative permeability} & \multicolumn{2}{c}{Brooks-Corey, $\lambda$ = 2}\\
			\multirow{4}*{Boundary conditions} & \multicolumn{2}{c}{$S_w(0,y,t)$ = 1, $p_n(0,y,t)=2\times10^{5}$ Pa}\\
			& \multicolumn{2}{c}{$S_w(300,y,t)$ = 0, $\phi_n(300,y,t)=3\times10^{-4}$ kg/(m$^2$s)}\\
			& \multicolumn{2}{c}{$\phi_w(x,0,t)=\phi_n(x,0,t)=0$ kg/(m$^2$s)}\\
			& \multicolumn{2}{c}{$\phi_w(x,75,t)=\phi_n(x,75,t)=0$ kg/(m$^2$s)}\\
			{Initial conditions} & \multicolumn{2}{c}{$S_w(x,y,0)$ = 0}\\
			\bottomrule
		\end{tabular}
		\label{case1}
	\end{center}
\end{table}
\begin{figure}[H]
	\begin{center}
		\includegraphics[width=0.6\linewidth]{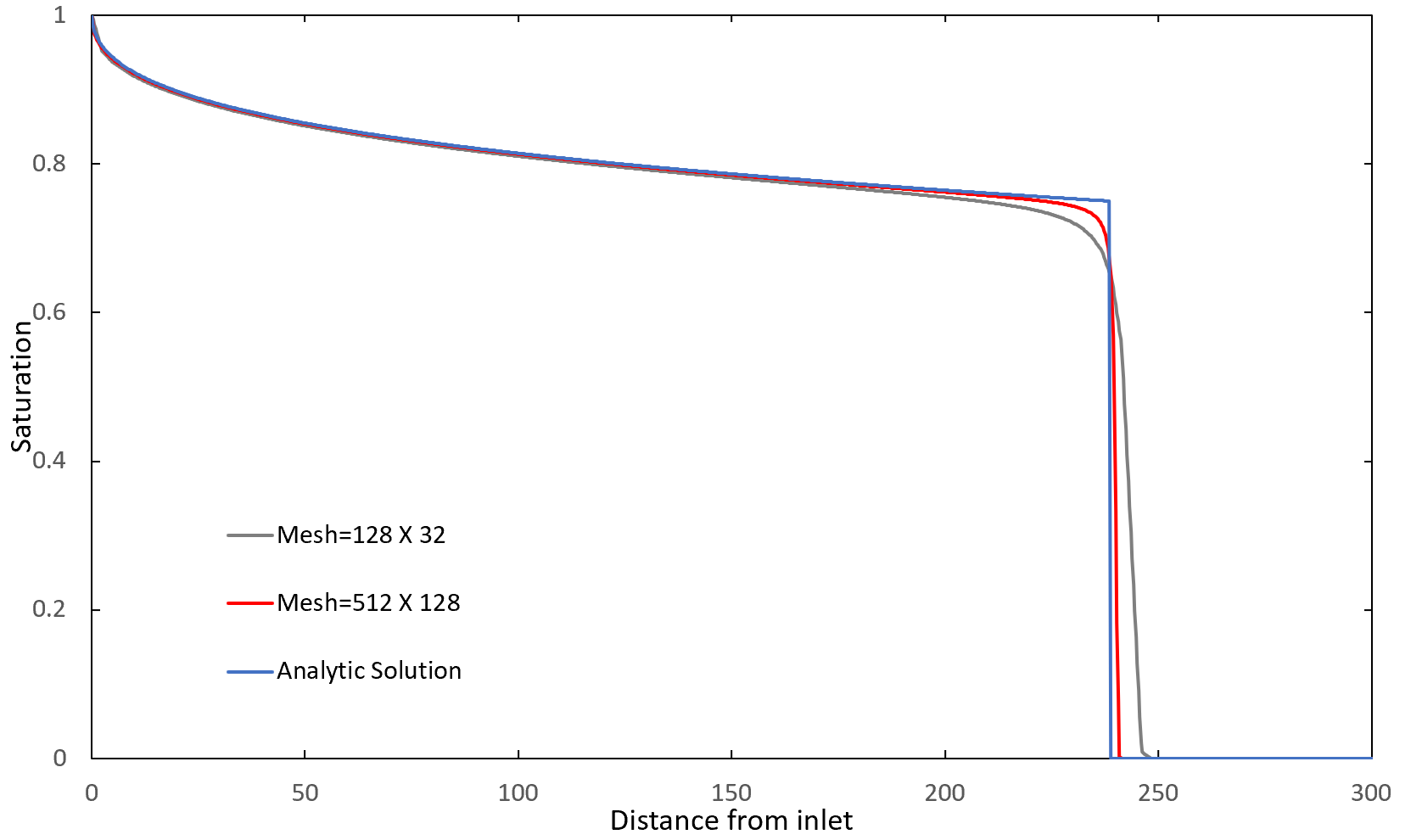}
		\caption{The profile curves of the wetting phase saturation with different mesh sizes for Case-\ref{case1} with Implicit Euler scheme}
		\label{BL}
	\end{center}
\end{figure}
\begin{figure}[H]
	\begin{center}
		\includegraphics[width=0.6\linewidth]{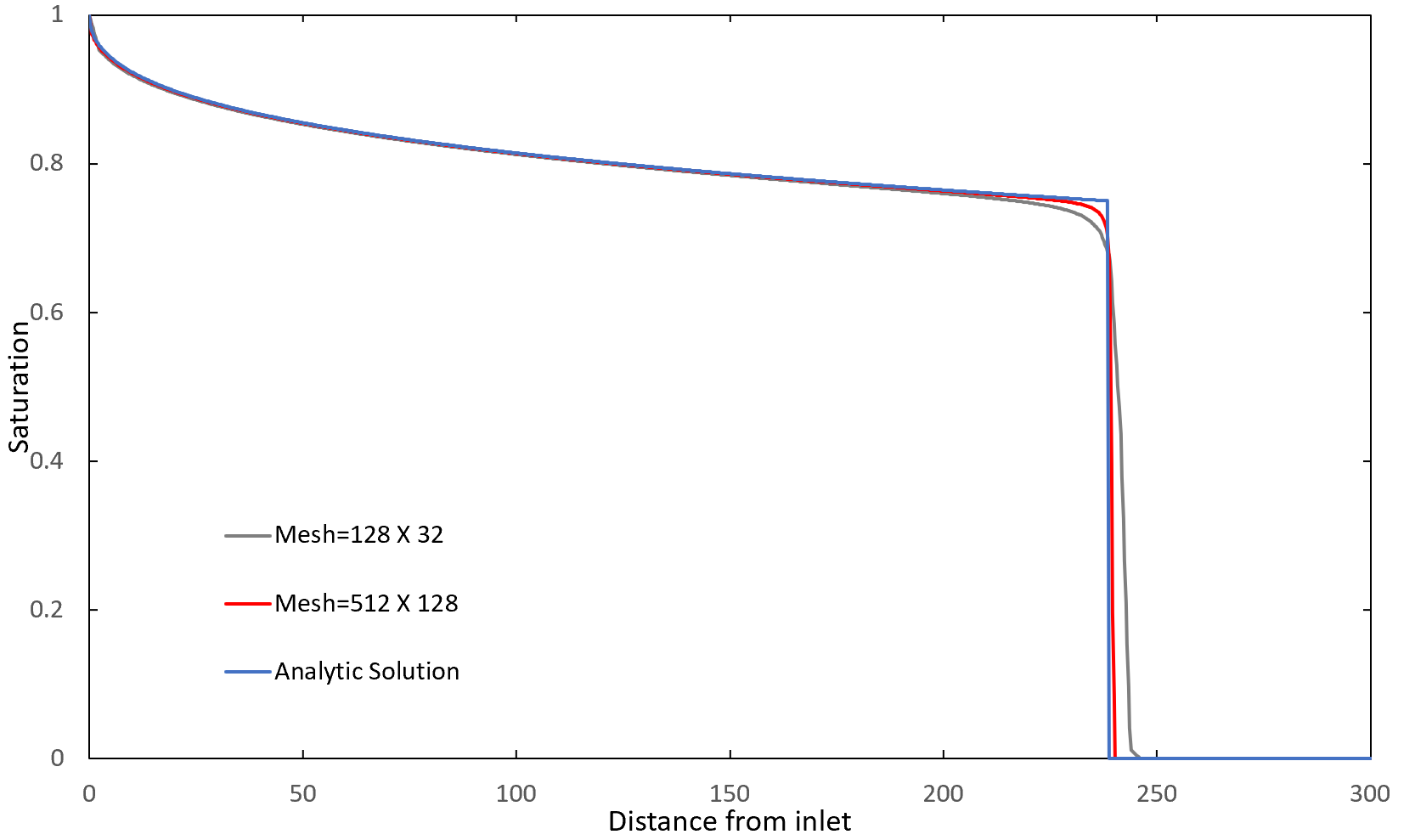}
		\caption{The profile curves of the wetting phase saturation with different mesh sizes for Case-\ref{case1} with LIMEX scheme}
		\label{BL_LIMEX}
	\end{center}
\end{figure}
With the parameters given in Table \ref{case1} the velocity of the front $v \approx 1.84\times10^{-6}$m/s. Courant number $C$ is set as 0.8 in Implicit Euler scheme. The fixed time step $\Delta t = C h/v $ and the final time $T=1500$ days. In the LIMEX scheme, the initial time step $\Delta t_0=1$ second and the maximum time step $\Delta t_{max}=100$ days. The error tolerance $\mathrm{TOL}$ and the safety factor $\rho$ is set as $5\times 10^{-2}$ and 0.25 respectively. The experiments are performed with 24 parallel processors.\\
The comparison of the saturation profiles of the wetting phase with different mesh sizes is shown in Figure \ref{BL} and Figure \ref{BL_LIMEX}. The shock fronts are described in the numerical solutions without oscillation. The sharp front is absent because of the smooth effect of the applied upwind scheme. The saturation profiles converge to the analytic solution with the mesh refinements.
To evaluate the convergence of the proposed numerical framework, the $L^p$-norm of the error and the convergence rate $r_{L^p}$ are defined as follows.
\begin{equation} \label{Lp}
	\left\|S_w-S_{w h}\right\|_{L^p}=\left(\int_{\Omega}\left|S_w-S_{w h}\right|^p d \mathbf{x}\right)^{\frac{1}{p}}
\end{equation}
\begin{equation} \label{r_Lp}
	r_{L^p}=\log_2 \left(\frac{\left\|S_w-S_{w 2 h}\right\|_{L^p}}{\left\|S_w-S_{w h}\right\|_{L^p}}\right)
\end{equation}
The error norms and convergence rates of the saturation at $T=1500$ days are shown in Table \ref{error1}. In Implicit Euler scheme, the time step is fined with the grid to keep a fixed Courant number. While in LIMEX scheme, the time step is selected adaptively. The accepted plus rejected time steps of LIMEX are counted in the table. The convergence of both time stepping schemes are great and the convergence rates $r_{L^2}$ are close to the theoretical value $1/2$.
\begin{table}[H]
	\begin{center}
		\caption{Error norms and convergence rates for Case-1}
		\begin{tabular}{*{7}{c}}
			\toprule
			{Time stepping scheme} & {Elements} & {Time steps} & {$L^1$ norm} & {$r_{L^1}$} & {$L^2$ norm}& {$r_{L^2}$}\\
			\midrule
			\multirow{4}*{Fixed(Implicit Euler)} & {64} & {65} & {8.739} & {-} & {1.663} & {-}\\
			& {128} & {130} & {4.951} & {0.820} & {1.250} & {0.411}\\
			& {256} & {260} & {2.766} & {0.840} & {0.931} & {0.425}\\
			& {512} & {520} & {1.521} & {0.862} & {0.685} & {0.442}\\
			\multirow{4}*{Adaptive(LIMEX)} & {64} & {483} & {5.797} & {-} & {1.425} & {-}\\
			& {128} & {654} & {3.225} & {0.846} & {1.060} & {0.427}\\
			& {256} & {860} & {1.768} & {0.867} & {0.777} & {0.448}\\
			& {512} & {1055} & {0.942} & {0.907} & {0.561} & {0.469}\\
			\bottomrule
		\end{tabular}
		\label{error1}
	\end{center}
\end{table}

\begin{figure}[H]
	\begin{center}
		\begin{tikzpicture}
			\draw[thick] (0,0) -- (0,2) -- (8,2)-- (8,0)-- (0,0);
			\draw[thick] (4,0) -- (4,2);
			\draw[] (2,1)  node[above]{$K_1$};
			\draw[] (6,1)  node[above]{$K_2$};
			\draw[] (2,1)  node[below]{$k_{rw,1}$, $k_{rn,1}$};
			\draw[] (6,1)  node[below]{$k_{rw,2}$, $k_{rn,2}$};
			
			\draw[|<->|,semithick]	(0,-0.2) -- (4,-0.2);
			\draw[<->|,semithick] 	(4,-0.2) -- (8,-0.2);
			\draw[] (2,-0.2)  node[below]{150m};
			\draw[] (6,-0.2)  node[below]{150m};
		\end{tikzpicture}
		\caption{Absolute and relative permeabilities for Case-2}
		\label{Arp}
	\end{center}
\end{figure}
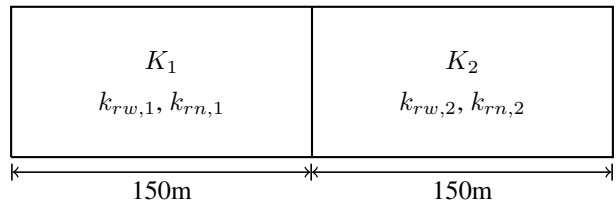

To test the performance of current framework in the simulation of immiscible two-phase flow in a heterogeneous porous medium, the extended Buckley-Leverett flow which is proposed by \cite{wu1993buckley}, is reproduced. The whole domain is divided into two parts with different absolute and relative permeabilities(Figure \ref{Arp}). The parameters for this test are provided in Table \ref{case2}. The error tolerance TOL in LIMEX is set as $10^{-2}$ to generate similar results to that of Implicit Euler.

\begin{table}[H]
	\begin{center}
		\caption{Parameters for Case-2}
		\begin{tabular}{*{3}{c}}
			\toprule
			\multicolumn{3}{c}{Case-2: Extended Buckley-Leverett}\\
			\midrule
			\multirow{2}*{Domain} & \multicolumn{2}{c}{300 m $\times$ 75m}\\
			& \multicolumn{2}{c}{domain1: $0\leq x\leq150$, domain2: $150<x\leq300$}\\
			{Rock properties} & \multicolumn{2}{c}{$\Phi=0.2$, $K_1=10^{-13}$ m$^2$, $K_2=10^{-14}$ m$^2$}\\
			\multirow{2}*{Fluid properties} & \multicolumn{2}{c}{$\rho_w=\rho_n=1\times10^3$ kg/m$^3$}\\
			& \multicolumn{2}{c}{$\mu_w=1\times10^{-3}$ Pa s,  $\mu_n=5\times10^{-3}$ Pa s}\\
			{Residual saturation} & \multicolumn{2}{c}{$S_{wr}=S_{nr}=0$, $S_e=\frac{S_w-S_{wr}}{1-S_{wr}-S_{nr}}$}\\
			{Capillary pressure} & \multicolumn{2}{c}{$p_c\equiv0$}\\
			\multirow{2}*{Relative permeability} & \multicolumn{2}{c}{$k_{rw,1}=1.831 S_e^4$, $k_{rn,1}=0.75 (1-1.25 S_e)^2(1-1.652 S_e^2)$}\\
			 & \multicolumn{2}{c}{$k_{rw,2}=0.4687 S_e^2$, $k_{rn,2}=0.25 (1-1.25 S_e)^2$}\\
			\multirow{4}*{Boundary conditions} & \multicolumn{2}{c}{$S_w(0,y,t)$ = 1, $p_n(0,y,t)=2\times10^{5}$ Pa}\\
			& \multicolumn{2}{c}{$S_w(300,y,t)$ = 0, $\phi_n(300,y,t)=2\times10^{-4}$ kg/(m$^2$s)}\\
			& \multicolumn{2}{c}{$\phi_w(x,0,t)=\phi_n(x,0,t)=0$ kg/(m$^2$s)}\\
			& \multicolumn{2}{c}{$\phi_w(x,75,t)=\phi_n(x,75,t)=0$ kg/(m$^2$s)}\\
			{Initial conditions} & \multicolumn{2}{c}{$S_w(x,y,0)$ = 0}\\
			\bottomrule
		\end{tabular}
		\label{case2}
	\end{center}
\end{table}

\begin{figure}[H]
	\begin{center}
		\includegraphics[width=0.6\linewidth]{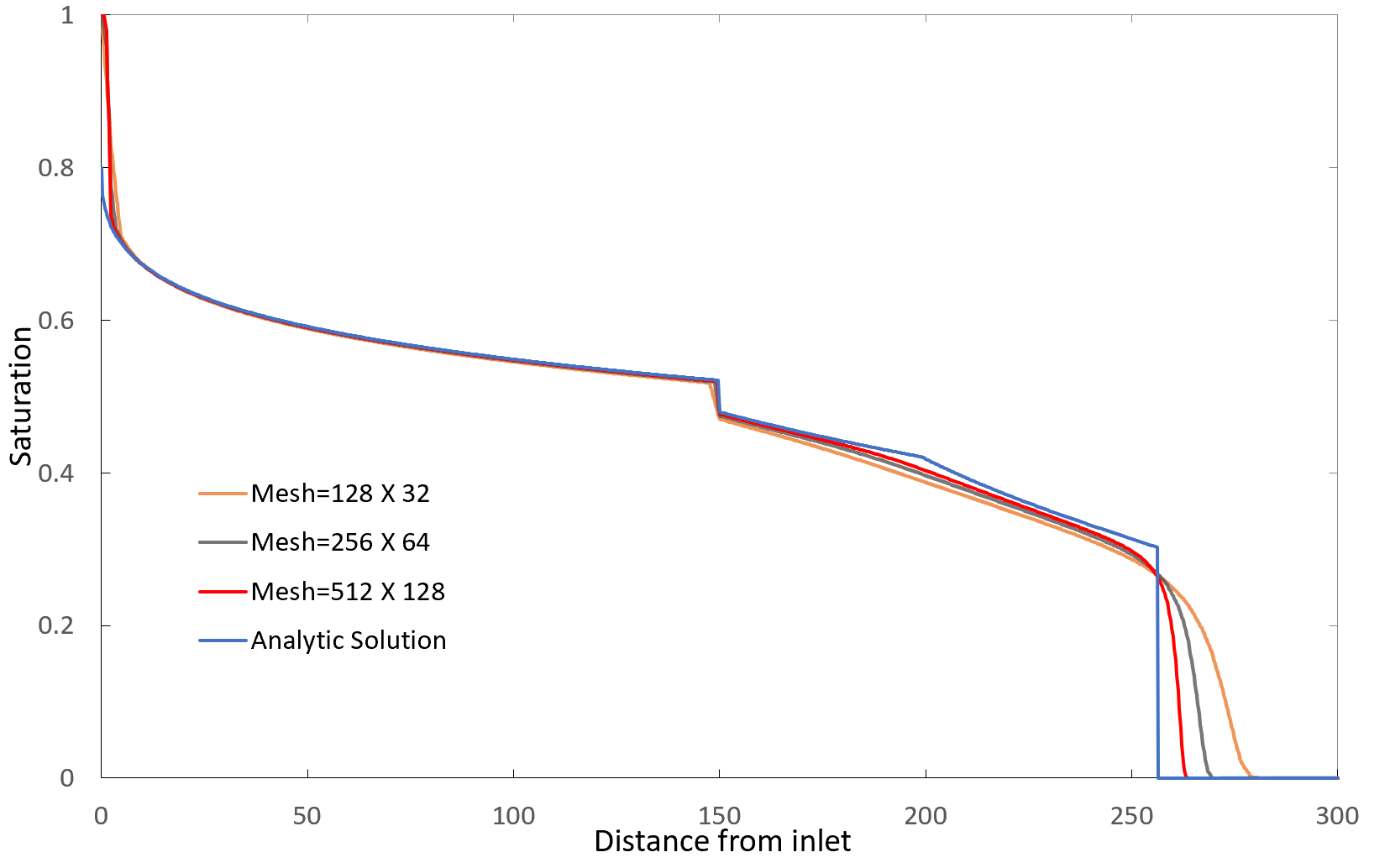}
		\caption{The profile curves of the wetting phase saturation with different mesh sizes for Case-2 with Implicit Euler scheme}
		\label{EBL}
	\end{center}
\end{figure}

\begin{figure}[H]
	\begin{center}
		\includegraphics[width=0.6\linewidth]{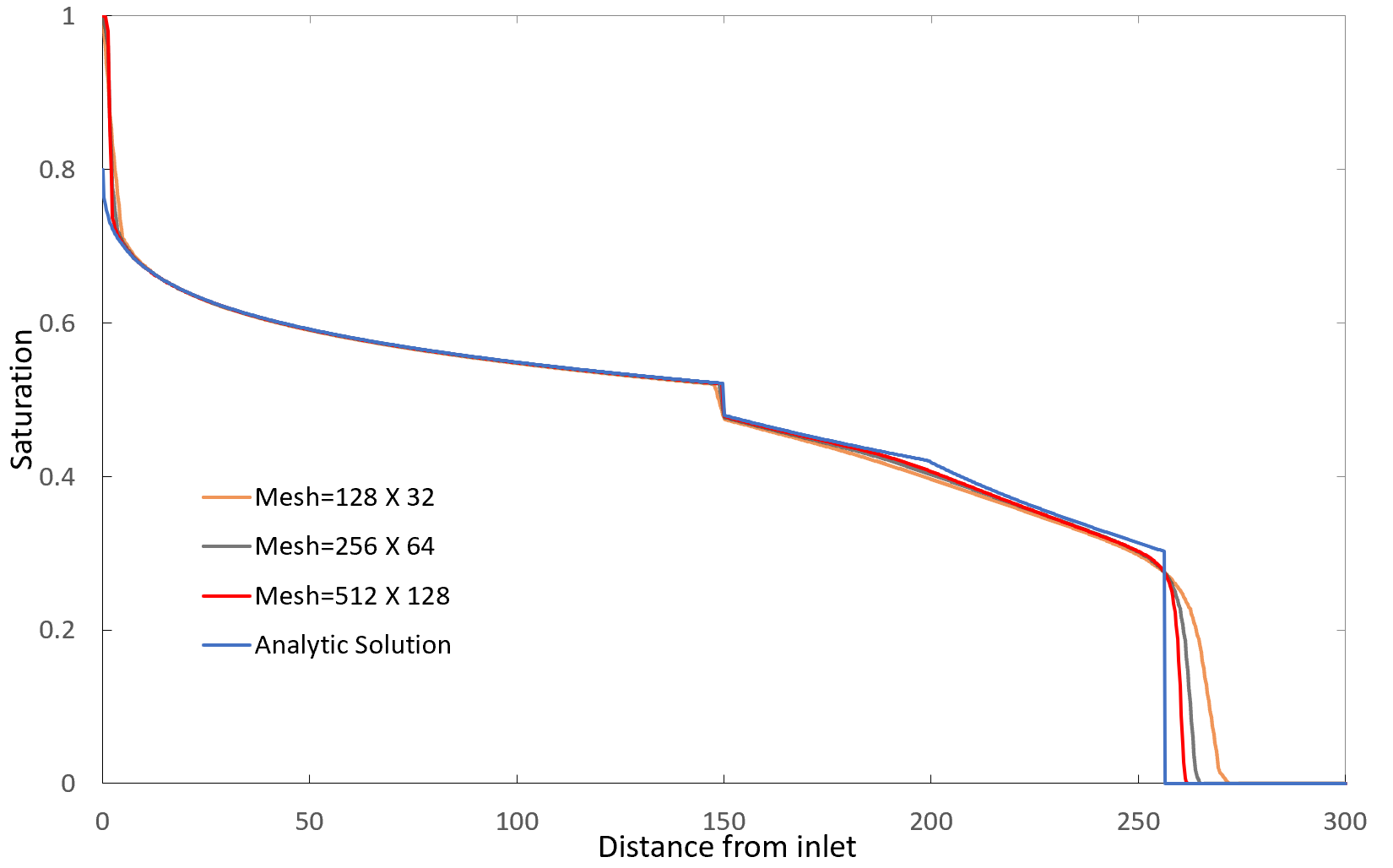}
		\caption{The profile curves of the wetting phase saturation with different mesh sizes for Case-2 with LIMEX scheme}
		\label{EBL_LIMEX}
	\end{center}
\end{figure}

The wetting phase saturation profiles with different mesh sizes are shown in Figure \ref{EBL} and \ref{EBL_LIMEX}. Due to the change in permeability, the saturation at the interface of the porous media(150 m from the inlet) jumped from 0.52 to 0.48. It is obvious that this discontinuity is captured by the proposed framework and the results converge to the analytic solution as the mesh is refined. The error norms and convergence rates of the saturation at $T=1500$ days are shown in Table \ref{error2}.

\begin{table}[H]
	\begin{center}
		\caption{Error norms and convergence rates for Case-2}
		\begin{tabular}{*{7}{c}}
			\toprule
			{Time stepping scheme} & {Elements} & {Time steps} & {$L^1$ norm} & {$r_{L^1}$} & {$L^2$ norm}& {$r_{L^2}$}\\
			\midrule
			\multirow{4}*{Fixed(Implicit Euler)} & {64} & {65} & {11.016} & {-} & {1.205} & {-}\\
			& {128} & {130} & {6.898} & {0.675} & {0.948} & {0.347}\\
			& {256} & {260} & {4.350} & {0.665} & {0.762} & {0.315}\\
			& {512} & {520} & {2.676} & {0.701} & {0.593} & {0.361}\\
			\multirow{4}*{Adaptive(LIMEX)} & {64} & {676+1} & {7.566} & {-} & {1.047} & {-}\\
			& {128} & {878+1} & {4.667} & {0.697} & {0.817} & {0.360}\\
			& {256} & {1195+1} & {2.931} & {0.671} & {0.657} & {0.315}\\
			& {512} & {1401+11} & {2.136} & {0.457} & {0.544} & {0.220}\\
			\bottomrule
		\end{tabular}
		\label{error2}
	\end{center}
\end{table}

In the previous two cases, the capillarity is neglected, which causes Equation (\ref{sW}) to degenerate from parabolic into a hyperbolic equation. To test the effect of capillary, the Mc-Whorter problem \cite{mcwhorter1990exact} is performed. The parameters for this simulation are listed in Table \ref{case3}.
\begin{table}[H]
	\begin{center}
		\caption{Parameters for Case-3}
		\begin{tabular}{*{3}{c}}
			\toprule
			\multicolumn{3}{c}{Case-3: Mc-Whorter}\\
			\midrule
			{Domain} & \multicolumn{2}{c}{1.6 m $\times$ 1.6m}\\
			{Rock properties} & \multicolumn{2}{c}{$\Phi=0.3$, $K=10^{-10}$ m$^2$}\\
			\multirow{2}*{Fluid properties} & \multicolumn{2}{c}{$\rho_w=\rho_n=1\times10^3$ kg/m$^3$}\\
			& \multicolumn{2}{c}{$\mu_w=\mu_n=1\times10^{-3}$ Pa s}\\
			{Residual saturation} & \multicolumn{2}{c}{$S_{wr}=S_{nr}=0$}\\
			{Capillary pressure} & \multicolumn{2}{c}{Brooks-Corey, $\lambda$ = 2, $p_d=5000$ Pa}\\
			{Relative permeability} & \multicolumn{2}{c}{Brooks-Corey, $\lambda$ = 2}\\
			\multirow{4}*{Boundary conditions} & \multicolumn{2}{c}{$S_w(0,y,t)$ = 1, $p_n(0,y,t)=2\times10^{5}$ Pa}\\
			& \multicolumn{2}{c}{$S_w(1.6,y,t)$ = 0, $\phi_n(1.6,y,t)=0$ kg/(m$^2$s)}\\
			& \multicolumn{2}{c}{$\phi_w(x,0,t)=\phi_n(x,0,t)=0$ kg/(m$^2$s)}\\
			& \multicolumn{2}{c}{$\phi_w(x,75,t)=\phi_n(x,75,t)=0$ kg/(m$^2$s)}\\
			{Initial conditions} & \multicolumn{2}{c}{$S_w(x,y,0)$ = 0}\\
			\bottomrule
		\end{tabular}
		\label{case3}
	\end{center}
\end{table}

\begin{figure}[H]
	\begin{center}
		\includegraphics[width=0.6\linewidth]{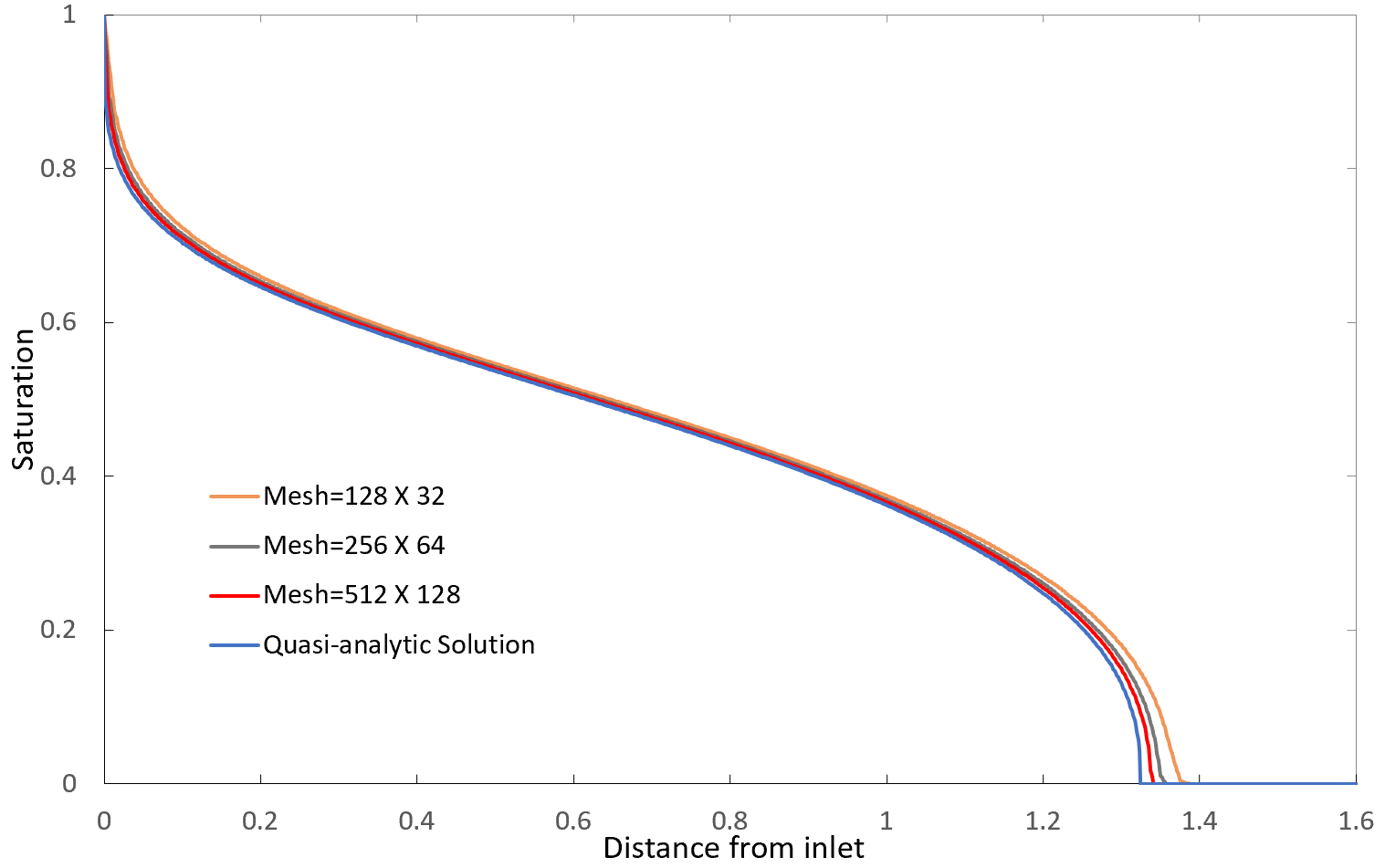}
		\caption{The profile curves of the wetting phase saturation with different mesh sizes for Case-3 with Implicit Euler scheme}
		\label{McW}
	\end{center}
\end{figure}

\begin{figure}[H]
	\begin{center}
		\includegraphics[width=0.6\linewidth]{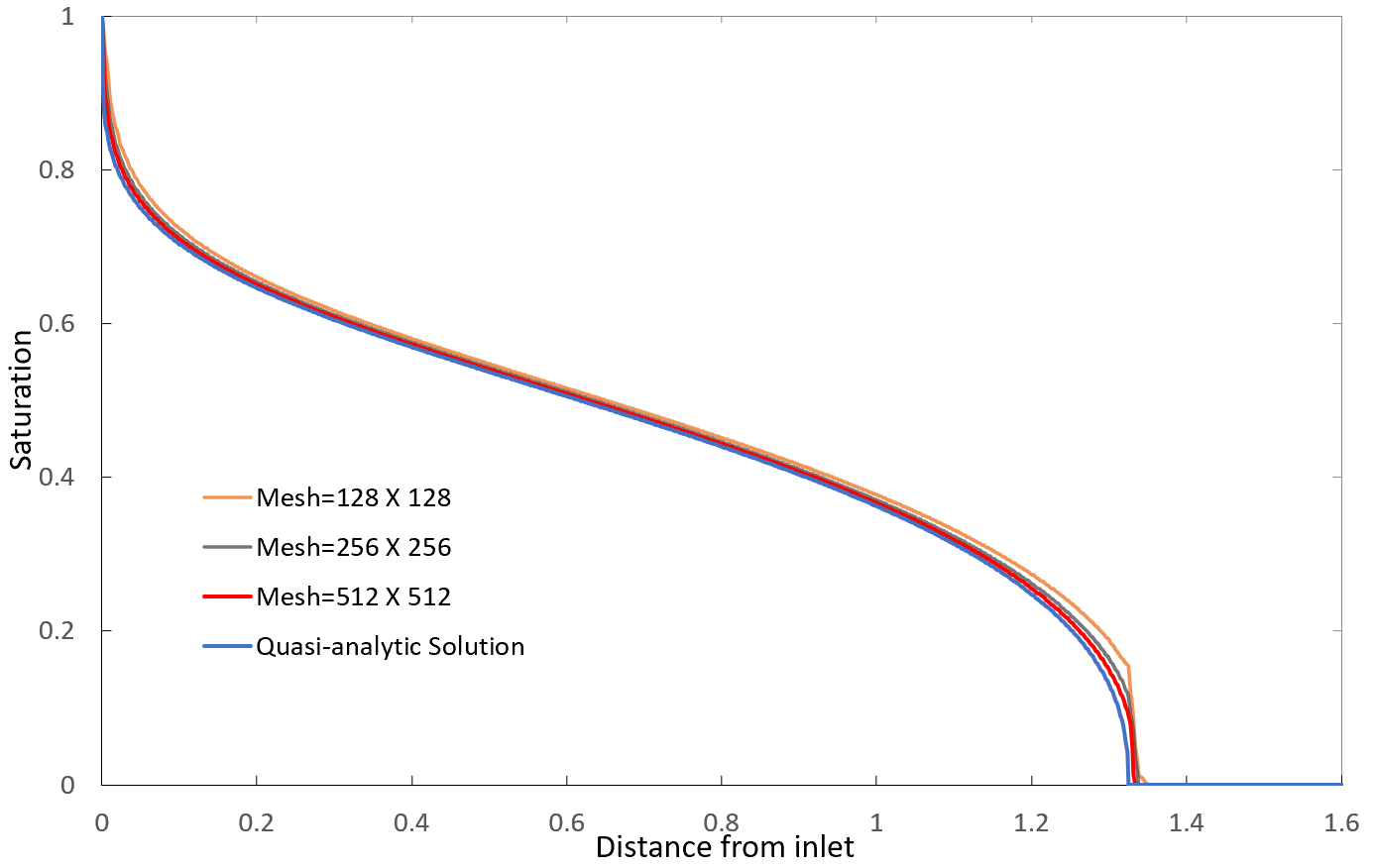}
		\caption{The profile curves of the wetting phase saturation with different mesh sizes for Case-3 with LIMEX scheme}
		\label{McW_LIMEX}
	\end{center}
\end{figure}

The domain is discretized with $64\times64$, $128\times128$, $256\times256$ and $512\times512$ elements. The number of fixed time steps are 24, 48, 96 and 192 respectively. In LIMEX scheme the error tolerance TOL is set as $8 \times 10^{-2}$ to generate similar results to that of Implicit Euler. The wetting phase saturation profiles at the final time $T=8000$ seconds with different mesh sizes are shown in Figure \ref{McW} and \ref{McW_LIMEX}. Compared to Buckley-Leverett flow, the solutions of Mc-Whorter flow are much smoother because of the capillary diffusion effect. It is obvious that the saturation profiles converge to the quasi-analytic solution as the mesh is refined. The error norms and convergence rates of the saturation at $T=8000$ seconds are shown in Table \ref{error3}.

\begin{table}[H]
	\begin{center}
		\caption{Error norms and convergence rates for Case-3}
		\begin{tabular}{*{7}{c}}
			\toprule
			{Time stepping scheme} & {Elements} & {Time steps} & {$L^1$ norm} & {$r_{L^1}$} & {$L^2$ norm}& {$r_{L^2}$}\\
			\midrule
			\multirow{4}*{Fixed(Implicit Euler)} & {64} & {24} & {0.045} & {-} & {0.051} & {-}\\
			& {128} & {48} & {0.026} & {0.816} & {0.032} & {0.681}\\
			& {256} & {96} & {0.014} & {0.836} & {0.020} & {0.704}\\
			& {512} & {192} & {0.008} & {0.853} & {0.012} & {0.726}\\
			\multirow{4}*{Adaptive(LIMEX)} & {64} & {180+2} & {0.045} & {-} & {0.047} & {-}\\
			& {128} & {209+3} & {0.025} & {0.848} & {0.028} & {0.722}\\
			& {256} & {327+3} & {0.014} & {0.830} & {0.017} & {0.695}\\
			& {512} & {619+3} & {0.008} & {0.842} & {0.011} & {0.691}\\
			\bottomrule
		\end{tabular}
		\label{error3}
	\end{center}
\end{table}

\begin{figure}[H]
	\begin{center}
		\begin{tikzpicture}
			\draw[thick] (0,0) -- (0,4) -- (10,4)-- (10,0)-- (0,0);
			\draw[thick] (0,1) -- (10,1);
			\draw[thick] (0,2) -- (10,2);
			\draw[thick] (0,3) -- (10,3);
			\draw[] (2.5,0.5)  node[]{$K_1$};
			\draw[] (2.5,1.5)  node[]{$K_2$};
			\draw[] (2.5,2.5)  node[]{$K_3$};
			\draw[] (2.5,3.5)  node[]{$K_4$};
			\draw[fill] (5,0.6) circle(0.6mm) node[below]{};
			
			\draw[|<->|,semithick]	(0,-0.2) -- (5,-0.2);
			\draw[<->|,semithick] 	(5,-0.2) -- (10,-0.2);
			\draw[] (2.5,-0.2)  node[below]{250m};
			\draw[] (7.5,-0.2)  node[below]{250m};
			
			\draw[|<->|,semithick]	(10.2,0) -- (10.2,0.6);
			\draw[<->|,semithick] 	(10.2,0.6) -- (10.2,4);
			\draw[] (10.2,0.3)  node[right]{30m};
			\draw[] (10.2,2.3)  node[right]{170m};
			
			\draw[dashed] (5,0.6) -- (5,0);
			\draw[dashed] (5,0.6) -- (10,0.6);
			\draw[] (5,0.5)  node[above]{Injection point};
		\end{tikzpicture}
		\caption{Absolute permeabilities and injection point for Case-4}
		\label{Ap}
	\end{center}
\end{figure}
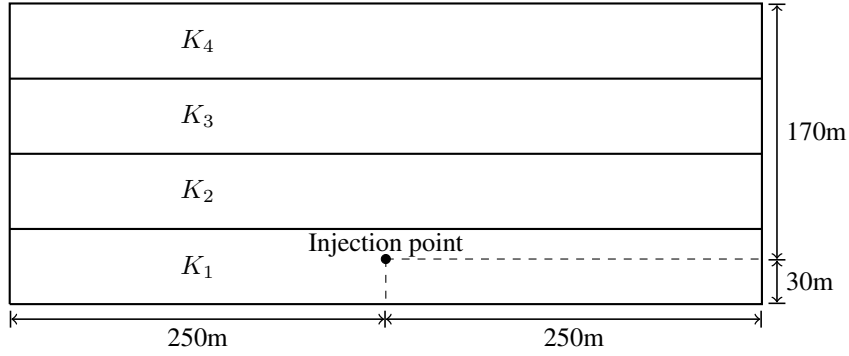

\begin{figure}[H]
	\begin{center}
		\includegraphics[width=0.6\linewidth]{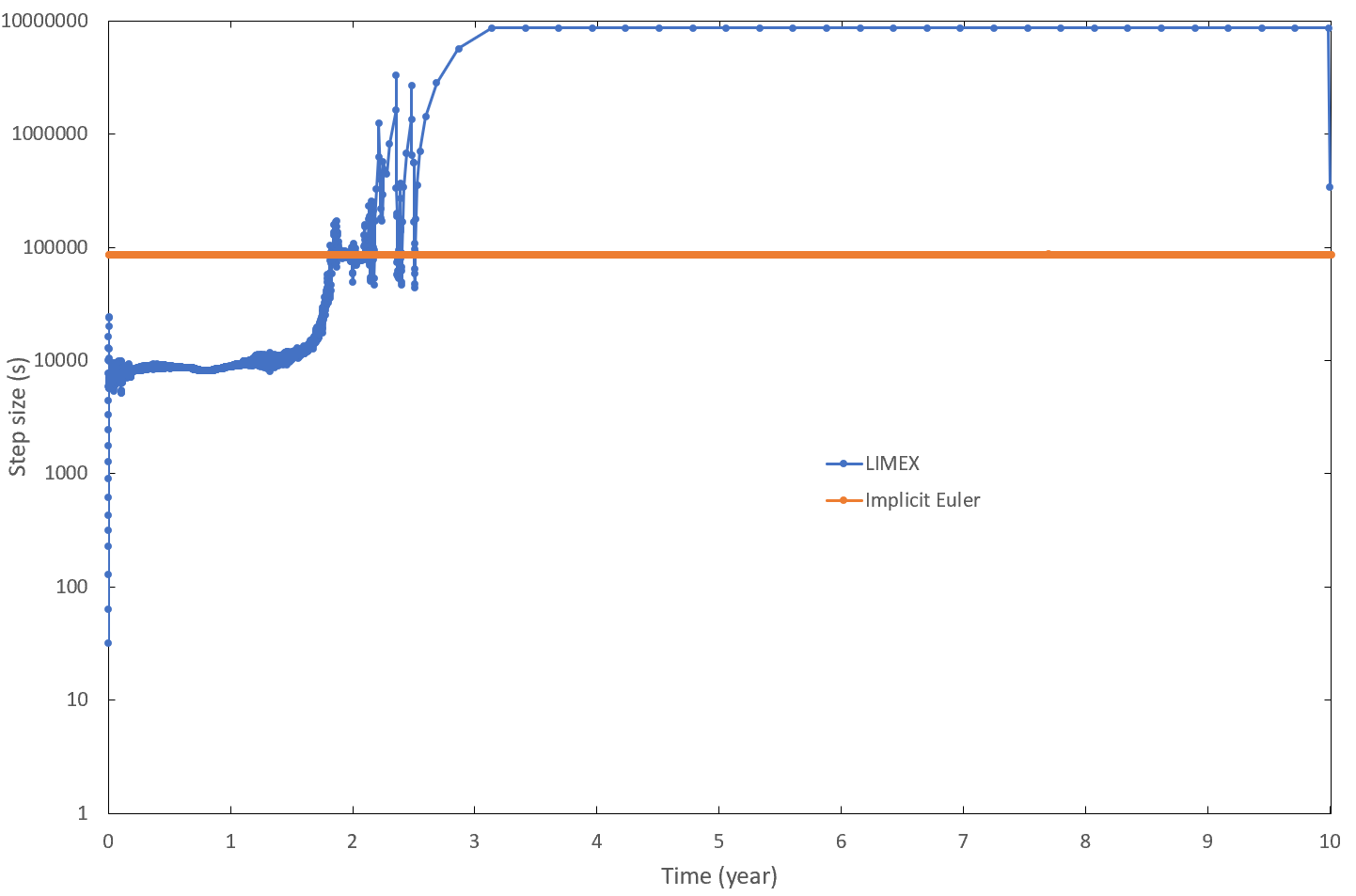}
		\caption{The time step size over the simulation time for Case-4}
		\label{StepSize}
	\end{center}
\end{figure}

To test the validation of the interface condition (Equation (\ref{hSw})) in heterogeneous porous medium, a layer-wise heterogeneity case is designed. The geometry consists of four layers with different absolute permeabilities, as shown in Figure \ref{Ap}. The parameters for this case are listed in Table \ref{case4}. 
\begin{table}[H]
	\begin{center}
		\caption{Parameters for Case-4}
		\begin{tabular}{*{3}{c}}
			\toprule
			\multicolumn{3}{c}{Case-4: 2D Layer-wise heterogeneity}\\
			\midrule
			{Domain} & \multicolumn{2}{c}{500 m $\times$ 200m}\\
			\multirow{2}*{Rock properties} & \multicolumn{2}{c}{$\Phi=0.1$, $K_1=10^{-13}$ m$^2$, $K_2=10^{-12}$ m$^2$}\\
			& \multicolumn{2}{c}{$K_3=8\times10^{-13}$ m$^2$, $K_4=7\times10^{-13}$ m$^2$}\\
			\multirow{2}*{Fluid properties} & \multicolumn{2}{c}{$\rho_w=1\times10^3$ kg/m$^3$, $\rho_n=7\times10^2$ kg/m$^3$}\\
			& \multicolumn{2}{c}{$\mu_w=1\times10^{-3}$ Pa s, $\mu_n=5.654\times10^{-5}$ Pa s}\\
			{Residual saturation} & \multicolumn{2}{c}{$S_{wr}=0.2$, $S_{nr}=0.1$, $S_e=\frac{S_w-S_{wr}}{1-S_{wr}-S_{nr}}$}\\
			{Capillary pressure} & \multicolumn{2}{c}{Brooks-Corey, $\lambda$ = 2, $p_d=7.37 \times K^{-0.43}$ Pa}\\
			{Relative permeability} & \multicolumn{2}{c}{$k_{rw}=0.9 S_e^2$, $k_{rn}=0.5 (1- S_e)$}\\
			{Injection rate} & \multicolumn{2}{c}{0.05 PV/year, for 2 years}\\
			\multirow{4}*{Boundary conditions} & \multicolumn{2}{c}{$\phi_w(0,y,t)=\phi_n(0,y,t)=0$ kg/(m$^2$s)}\\
			& \multicolumn{2}{c}{$\phi_w(500,y,t)=\phi_n(500,y,t)=0$ kg/(m$^2$s)}\\
			& \multicolumn{2}{c}{$S_w(x,0,t) = 1-S_{nr}$, $p_n(x,0,t)=5\times10^{7}$ Pa}\\
			& \multicolumn{2}{c}{$\phi_w(x,200,t)=\phi_n(x,200,t)=0$ kg/(m$^2$s)}\\
			{Initial conditions} & \multicolumn{2}{c}{$S_w(x,y,0)=1-S_{nr}$}\\
			\bottomrule
		\end{tabular}
		\label{case4}
	\end{center}
\end{table}
In the simulation of Case-4, the mesh size is $256\times128$. It is carried out on Shaheen III with 64 processors. Implicit Euler and LIMEX methods are employed for fixed and adaptive time stepping respectively. The fixed step size $\Delta t =1$ day for Implicit Euler scheme. In LIMEX scheme, the error tolerance TOL is set as $5 \times 10^{-3}$, the safety factor $\rho$ is 0.75, the initial step size is 32 seconds and the maximum step size is 100 days.\\
The time step sizes over the simulation time $T=10$ years are plotted in Figure \ref{StepSize}. In the beginning, the step size of LIMEX increased rapidly and varied around 9000 seconds. The step size increases again when the solution is steady until the injection stops (at $T=2$ years). At the beginning of the post-injection stage, the step size changes a lot since the solution state changes. After that, the step size increased again and reached the maximum step size. The comparison of fixed and adaptive time stepping for Case-4 is shown in Table \ref{case4_2}. It is obvious that LIMEX takes more time steps but there is no nonlinear iteration in each time step. Its total number of linear iterations is much less than that of Implicit Euler scheme. Therefore, the total execution time of LIMEX is much less than that of Implicit Euler, which makes it a great time stepping scheme for multi-phase simulations, especially over long periods.

\begin{table}[H]
	\begin{center}
		\caption{The comparison of the fixed and adaptive time stepping scheme for Case-4}
		\begin{tabular}{*{5}{c}}
			\toprule
			{Time stepping scheme} & {$T_{\text{total}}$ (s)} & {$N_{\text{time step}}$} & {$N_{\text{nonlinear}}$} & {$N_{\text{linear}}$}\\
			\midrule
			{Fixed(Implicit Euler)} & {3938.76} & {3650} & {176596} & {375093}\\
			{Adaptive(LIMEX)} & {275.28} & {6206+75} & {-} & {58093}\\
			\bottomrule
		\end{tabular}
		\label{case4_2}
	\end{center}
\end{table}

\begin{figure}[htbp]
	\centering
	\subfloat[$T$ = 36 days, Implicit Euler]
	{\includegraphics[width=0.4\textwidth]{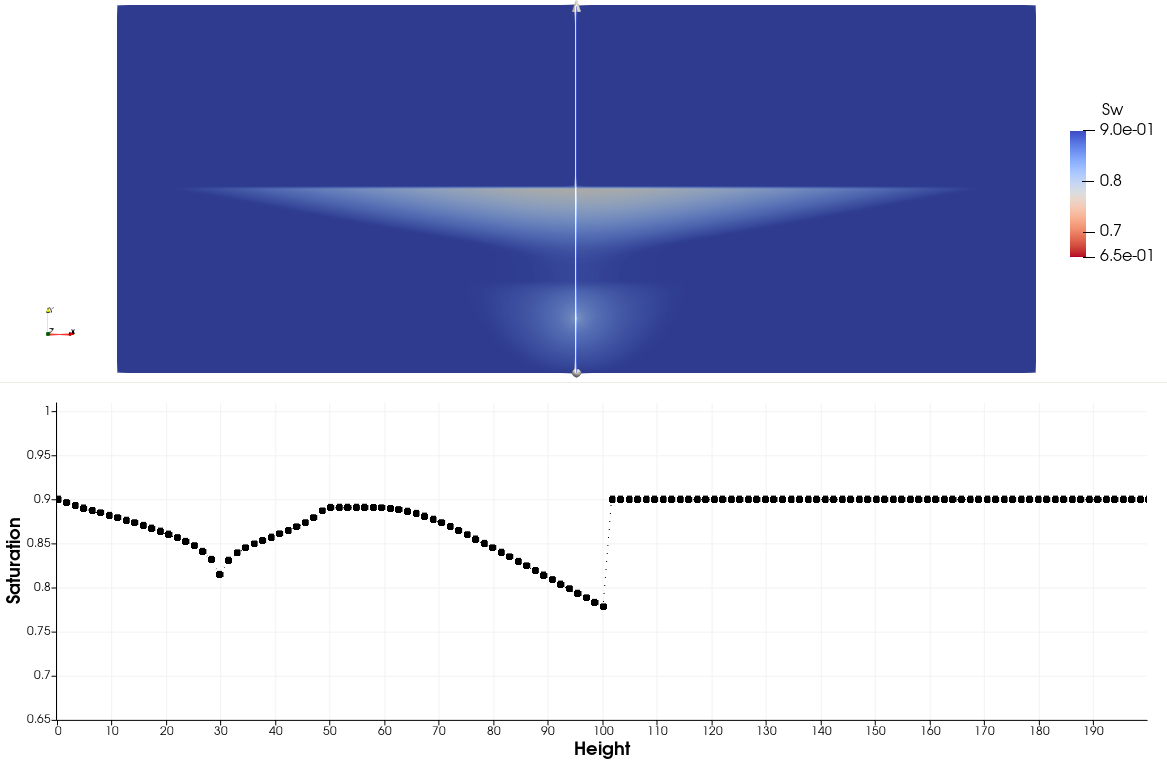}\label{T35Implicit}}\quad
	\subfloat[$T$ = 36 days, LIMEX]
	{\includegraphics[width=0.4\textwidth]{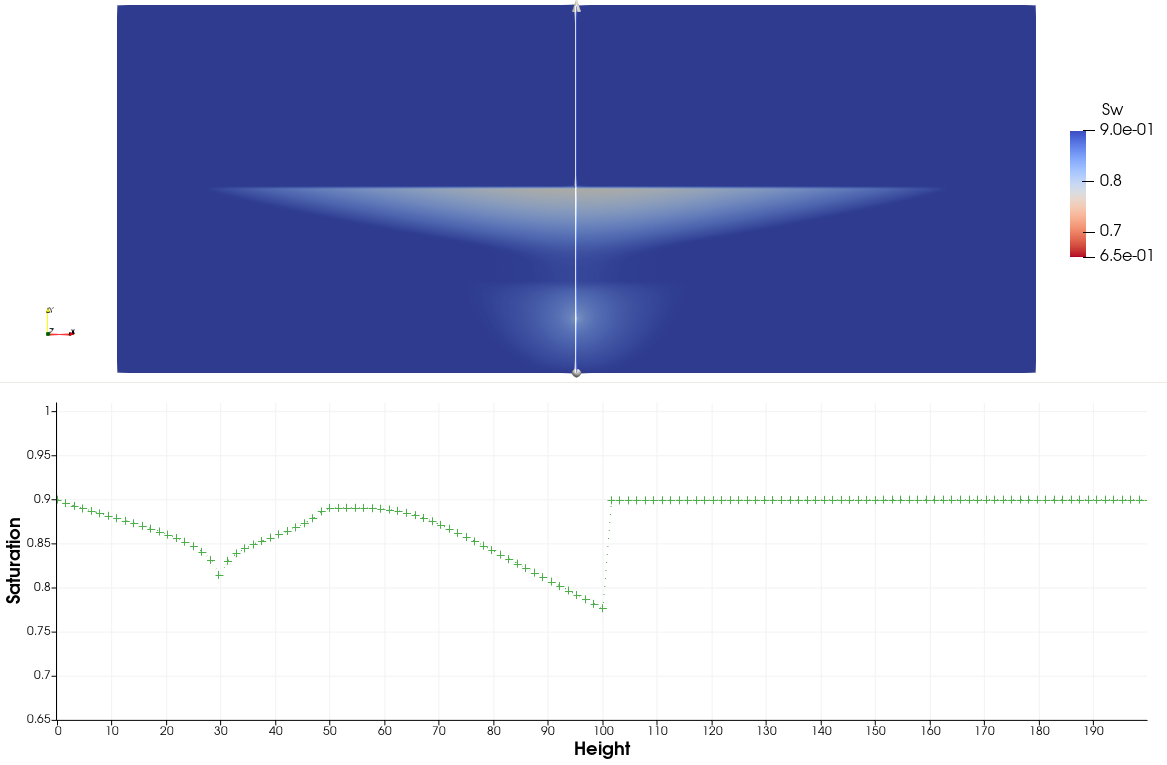}\label{T35LIMEX}}

	\subfloat[$T$ = 41 days, Implicit Euler]
	{\includegraphics[width=0.4\textwidth]{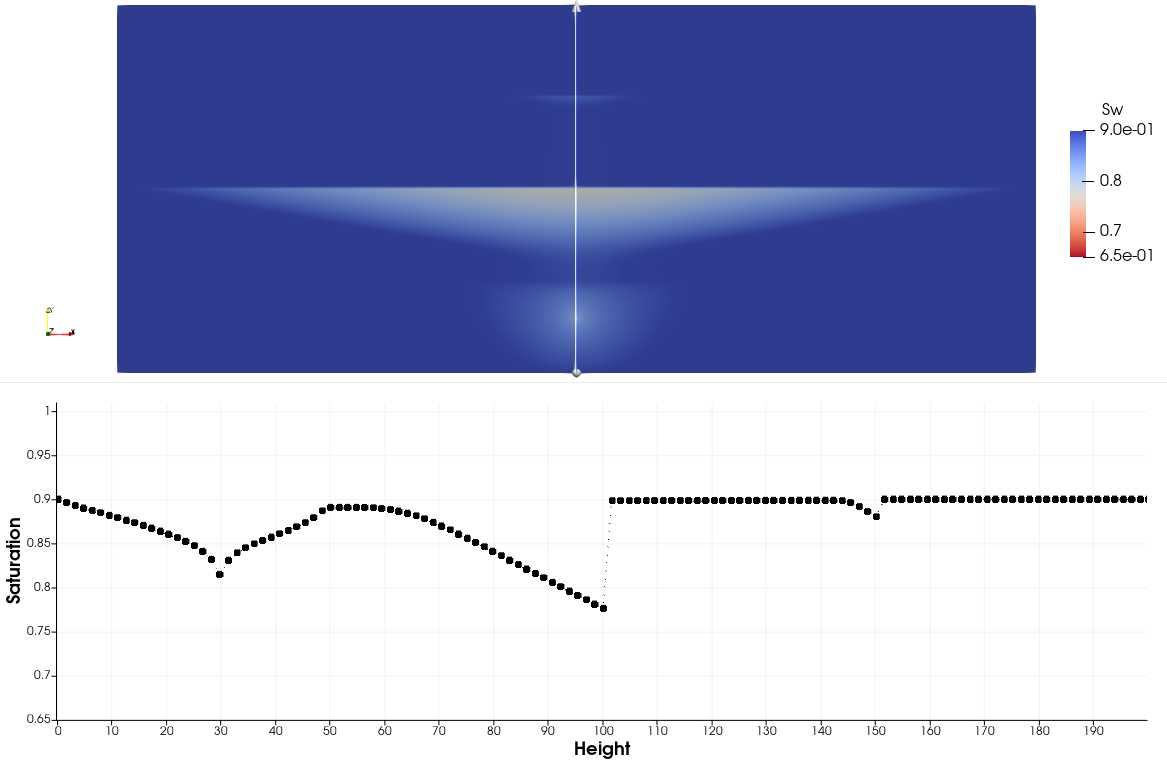}\label{T40Implicit}}\quad
	\subfloat[$T$ = 41 days, LIMEX]
	{\includegraphics[width=0.4\textwidth]{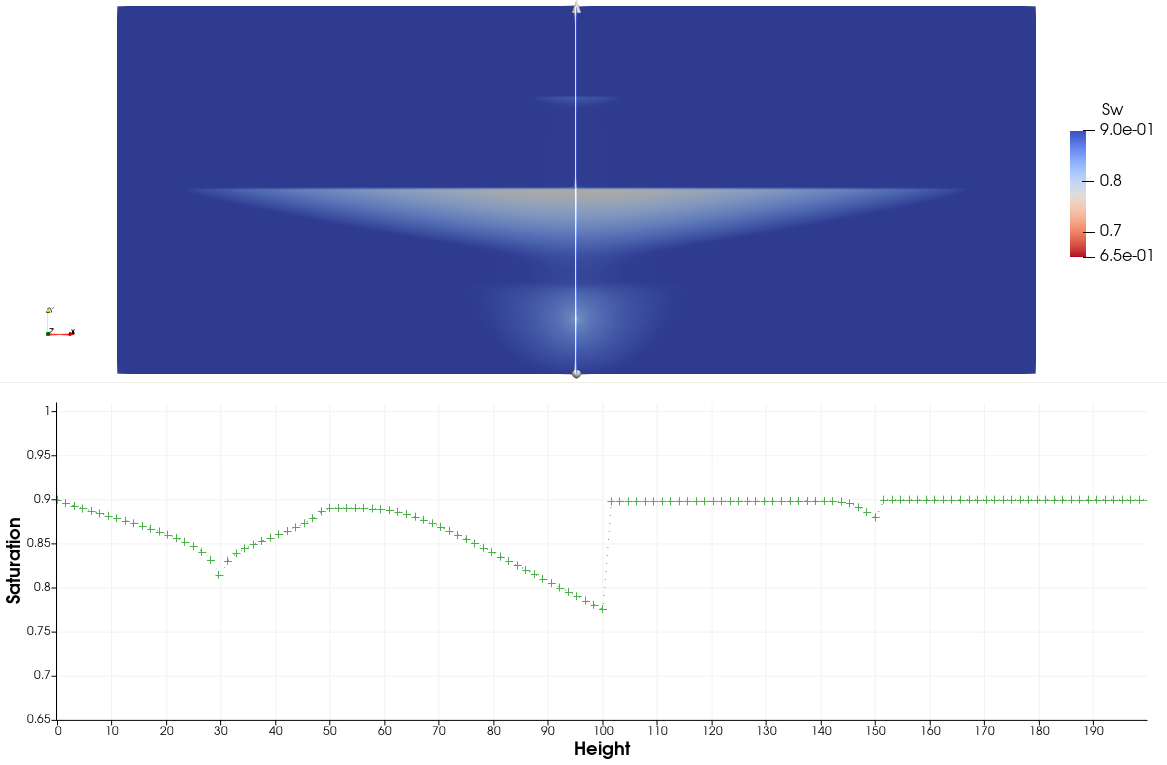}\label{T40LIMEX}}

	\subfloat[$T$ = 61 days, Implicit Euler]
	{\includegraphics[width=0.4\textwidth]{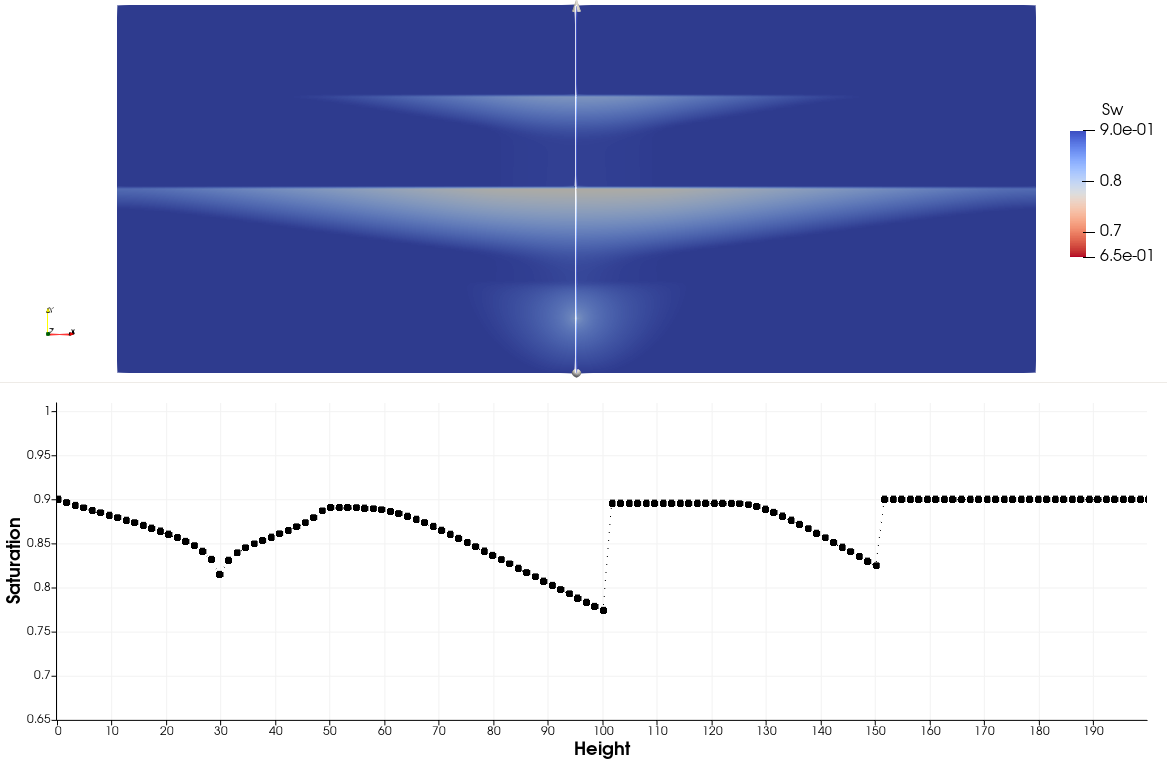}\label{T60Implicit}}\quad
	\subfloat[$T$ = 61 days, LIMEX]
	{\includegraphics[width=0.4\textwidth]{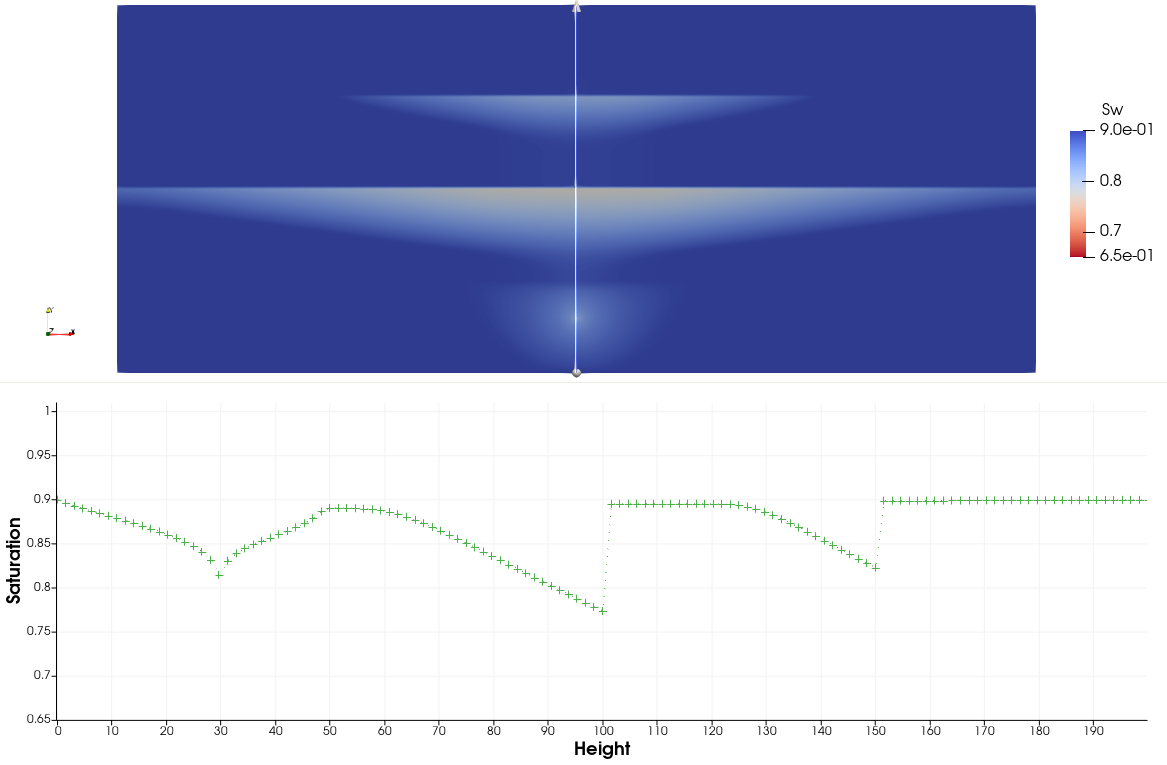}\label{T60LIMEX}}
	
	\subfloat[$T$ = 81 days, Implicit Euler]
	{\includegraphics[width=0.4\textwidth]{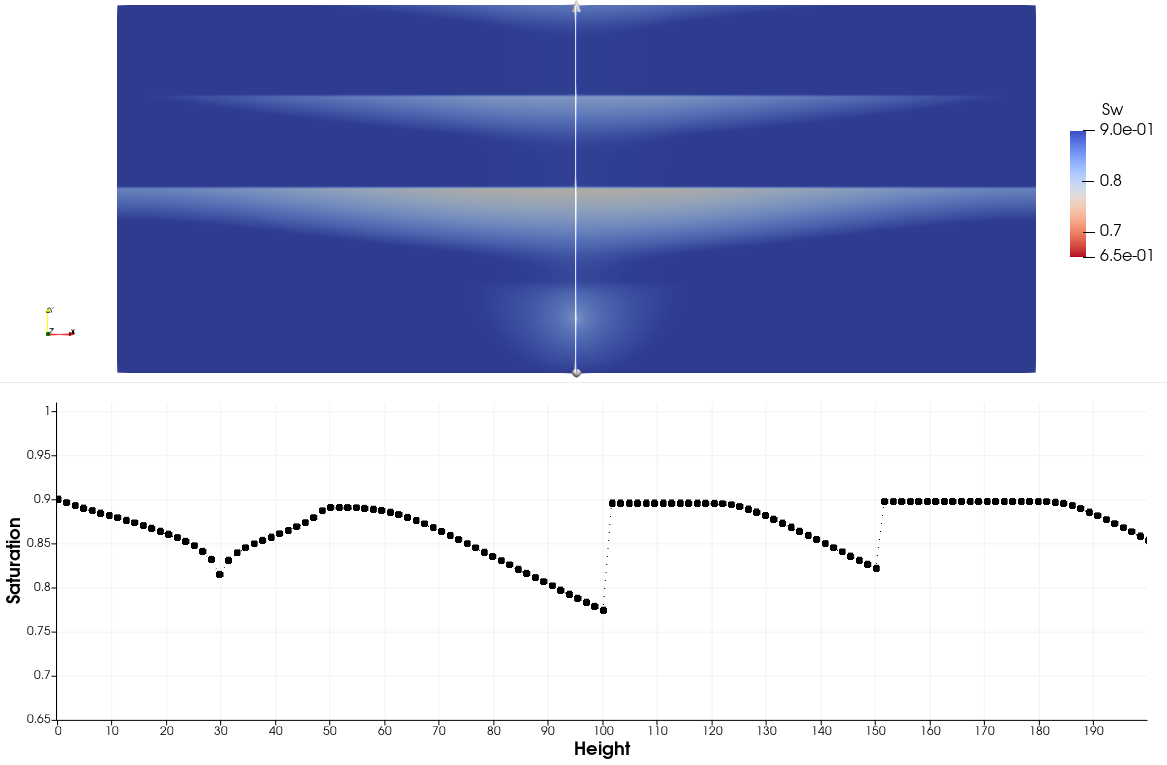}\label{T80Implicit}}\quad
	\subfloat[$T$ = 81 days, LIMEX]
	{\includegraphics[width=0.4\textwidth]{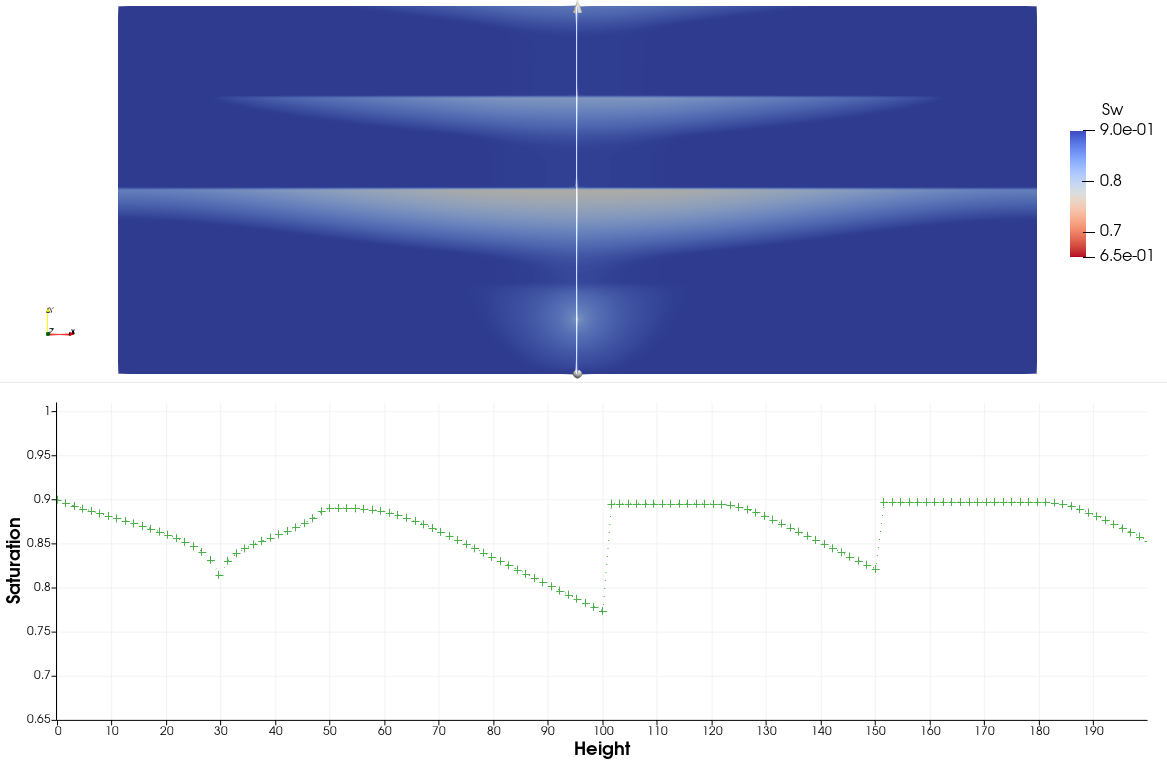}\label{T80LIMEX}}

	\caption{Contour of wetting-phase saturation and the distribution along central axis with fixed and adaptive time stepping for Case-4}
	\label{contour}
\end{figure}

\begin{figure}[htbp]
	\centering	
	\subfloat[$T$ = 3650 days, Implicit Euler]
	{\includegraphics[width=0.4\textwidth]{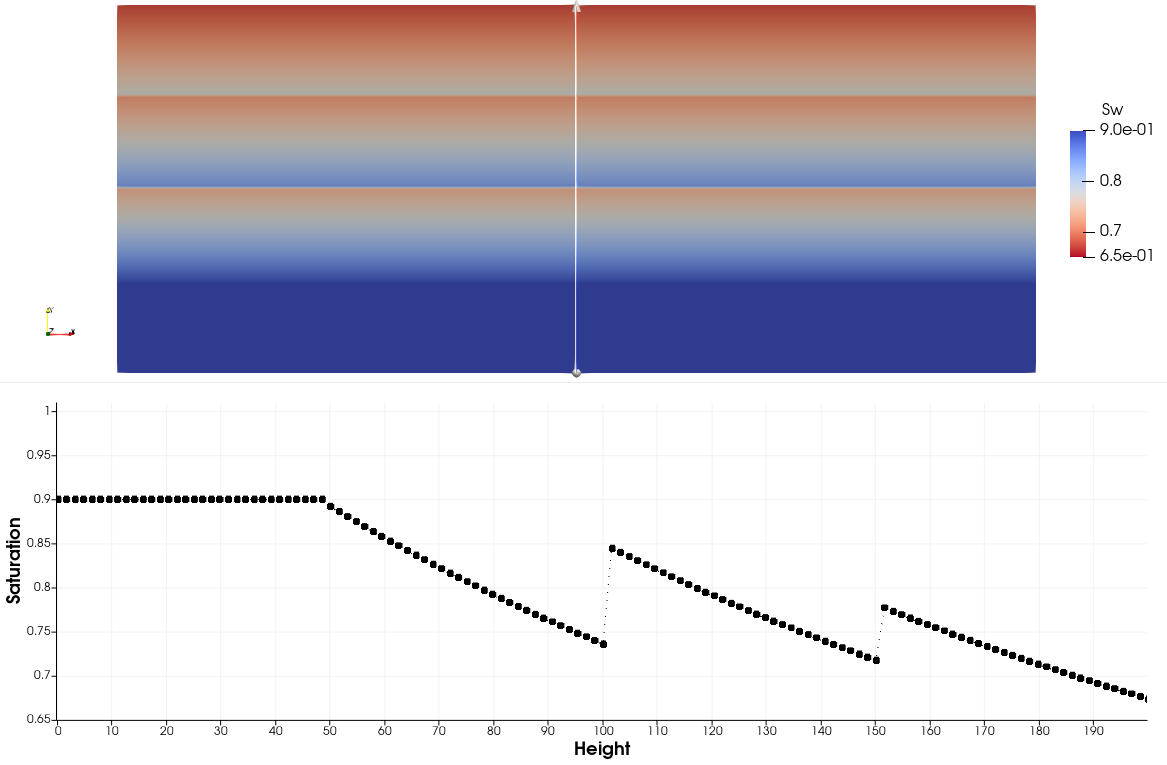}\label{T3649Implicit}}\quad
	\subfloat[$T$ = 3650 days, LIMEX]
	{\includegraphics[width=0.4\textwidth]{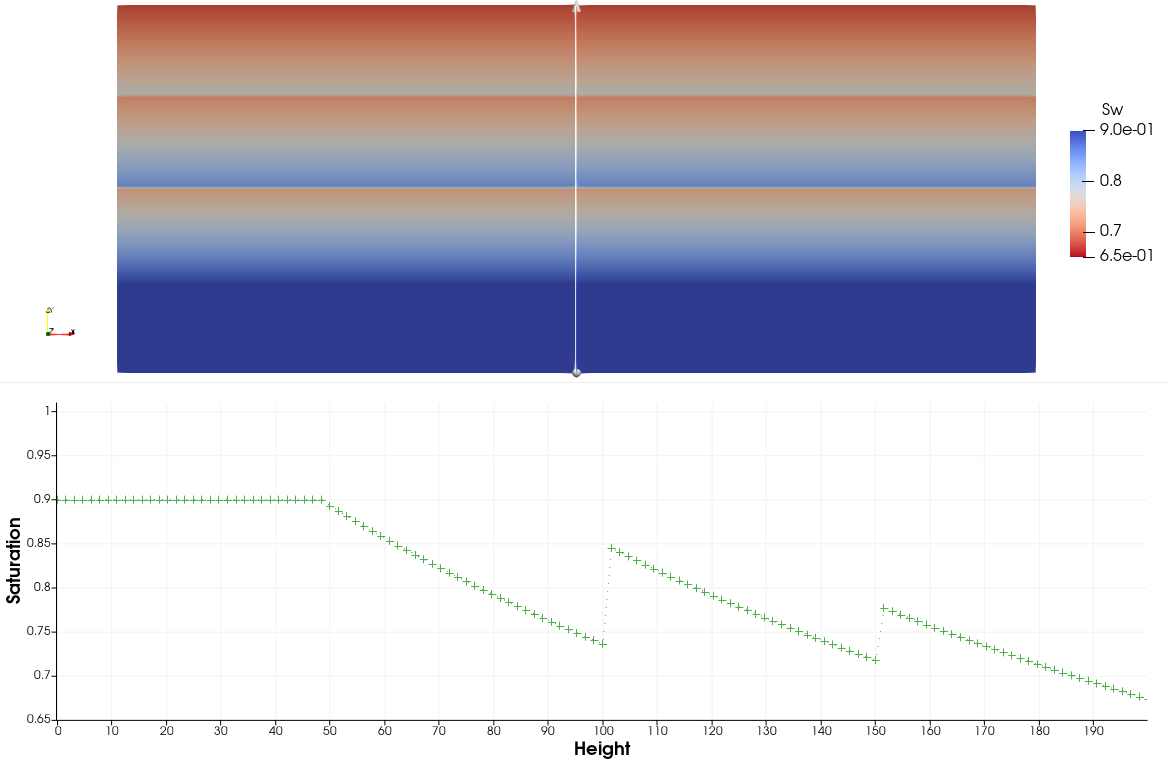}\label{T3649LIMEX}}
	
	\caption{Contour of wetting-phase saturation and the distribution along central axis with fixed and adaptive time stepping for Case-4}
	\label{contour2}
\end{figure}

The contour of wetting-phase saturation and the distribution along the central axis are shown in Figure \ref{contour} and Figure \ref{contour2}. In Figure \ref{contour}(a,b), non-wetting phase was injected from the first layer and floating up to the second layer due to buoyancy. Since the entry pressure of the third layer is higher than that of the second layer, non-wetting phase was trapped in the second layer until $S_w$ is lower than the threshold of wetting phase saturation $S_w^*$. With the parameters given in Table \ref{case4} one can get $S_w^*$ at the interface between the second and the third layers is 0.7778. In Figure \ref{contour}(c,d), $S_w$ at the interface between the second and the third layers decreased and it was lower than $S_w^*$, the non-wetting phase entered the third layer. Similarly, non-wetting phase was trapped in the third layer for a while (Figure \ref{contour}(e,f)) and broke through with the accumulation of non-wetting phase on the top layer (Figure \ref{contour}(g,h)). After injection, most of non-wetting phase entered the third and the fourth layer through the second layer. However, $S_w$ in the second layer will increase with non-wetting phase leaving. At some stage, the non-wetting phase will be trapped again because $S_w$ is not low enough anymore. Similar phenomena happened in the third layer. Finally, most of non-wetting phase is accumulated in the fourth layer and the rest is trapped in the second and the third layers(Figure \ref{contour2}).
It is clear that the interface condition in the current framework is able to capture saturation discontinuities in the simulation of multi-phase flow in heterogeneous porous media.

To evaluate the parallel performance of the proposed framework, the strong and weak scaling tests are performed for Case-4. The simulation time $T=128$ mins. The fixed step size $\Delta t =128$ seconds for Implicit Euler scheme. In LIMEX scheme, the initial step size is 32 seconds, with a maximum step size of 100 days. The results of the strong scaling test are presented in Table \ref{case4_strong}. The mesh size is fixed as $4096\times2048$ with 16,789,506 DoFs in the test. Notably, both the total execution time $T_{\text{total}}$ and the time required for each linear iteration $t_{\text{linear}}$ exhibit favorable scaling trends with the increasing number of processors.\\

\begin{table}[H]
	\begin{center}
		\caption{Strong scaling test for Case-4}
		\begin{tabular}{*{9}{c}}
		\toprule
		{} & {PE} & {$T_{\text{total}}$(s)} & {$N_{\text{time step}}$} & {$N_{\text{nonlinear}}$} & {$t_{\text{nonlinear}}$(s)} & {$N_{\text{linear}}$} & {$t_{\text{linear}}$(s)} & {Speedup}\\
		\midrule
		\multirow{4}*{Implicit Euler} & {64} & {1314.9} & {60} & {512} & {2.57} & {709} & {0.292} & {1}\\
		& {128} & {707.9} & {60} & {512} & {1.38} & {709} & {0.142} & {1.857}\\
		& {256} & {359.5} & {60} & {512} & {0.70} & {709} & {0.068} & {1.969}\\
		& {512} & {182.4} & {60} & {512} & {0.36} & {727} & {0.033} & {1.971}\\
		\midrule
		\multirow{4}*{LIMEX} & {64} & {487.4} & {47+1} & {-} & {-} & {307} & {0.290} & {1}\\
		& {128} & {267.3} & {47+1} & {-} & {-} & {307} & {0.140} & {1.823}\\
		& {256} & {130.7} & {47+1} & {-} & {-} & {307} & {0.067} & {2.045}\\
		& {512} & {65.0} & {47+1} & {-} & {-} & {311} & {0.033} & {2.011}\\
		\bottomrule
		\end{tabular}
		\label{case4_strong}
	\end{center}
\end{table}

\begin{table}[H]
	\begin{center}
		\caption{Weak scaling test for Case-4}
		\begin{tabular}{*{9}{c}}
			\toprule
			{} & {PE} & {DoF} & {$T_{\text{total}}$(s)} & {$N_{\text{time step}}$} & {$N_{\text{nonlinear}}$} & {$t_{\text{nonlinear}}$(s)} & {$N_{\text{linear}}$} &  {$t_{\text{linear}}$(s)}\\
			\midrule
			\multirow{4}*{Implicit Euler} & {64} & {16,789,506} & {1314.9} & {60} & {512} & {2.57} & {709} & {0.292}\\
			& {256} & {67,133,442} & {2385.2} & {120} & {854} & {2.79} & {1146} & {0.299}\\
			& {1024} & {268,484,610} & {4457.3} & {240} & {1428} & {3.12} & {1811} & {0.332}\\
			& {4096} & {1,073,840,130} & {7334.6} & {480} & {2342} & {3.13} & {2903} & {0.316}\\
			\midrule
			\multirow{4}*{LIMEX} & {64} & {16,789,506} & {487.4} & {47+1} & {-} & {-} & {307} & {0.290}\\
			& {256} & {67,133,442} & {802.9} & {73+1} & {-} & {-} & {438} & {0.297}\\
			& {1024} & {268,484,610} & {1698.1} & {133+6} & {-} & {-} & {817} & {0.327}\\
			& {4096} & {1,073,840,130} & {3228.8} & {254+23} & {-} & {-} & {1538} & {0.309}\\
			\bottomrule
		\end{tabular}
		\label{case4_weak}
	\end{center}
\end{table}

The last case is similar to Case-4 but has been extended from 2D to 3D. Here we use this 3D heterogeneous case to test the parallel performance of LIMEX stepping scheme in the proposed framework. The parameters for this test case are listed in Table \ref{case5}. The simulation time $T=128$ mins. The initial step size is 32 seconds and the maximum step size is 100 days.
\begin{table}[H]
	\begin{center}
		\caption{Parameters for Case-5}
		\begin{tabular}{*{3}{c}}
			\toprule
			\multicolumn{3}{c}{Case-5: 3D Layer-wise heterogeneity}\\
			\midrule
			{Domain} & \multicolumn{2}{c}{200 m $\times$ 200m $\times$ 200m}\\
			\multirow{2}*{Rock properties} & \multicolumn{2}{c}{$\Phi=0.1$, $K_1=10^{-13}$ m$^2$, $K_2=10^{-12}$ m$^2$}\\
			& \multicolumn{2}{c}{$K_3=8\times10^{-13}$ m$^2$, $K_4=7\times10^{-13}$ m$^2$}\\
			\multirow{2}*{Fluid properties} & \multicolumn{2}{c}{$\rho_w=1\times10^3$ kg/m$^3$, $\rho_n=7\times10^2$ kg/m$^3$}\\
			& \multicolumn{2}{c}{$\mu_w=1\times10^{-3}$ Pa s, $\mu_n=5.654\times10^{-5}$ Pa s}\\
			{Residual saturation} & \multicolumn{2}{c}{$S_{wr}=0.2$, $S_{nr}=0.1$, $S_e=\frac{S_w-S_{wr}}{1-S_{wr}-S_{nr}}$}\\
			{Capillary pressure} & \multicolumn{2}{c}{Brooks-Corey, $\lambda$ = 2, $p_d=7.37 \times K^{-0.43}$ Pa}\\
			{Relative permeability} & \multicolumn{2}{c}{$k_{rw}=0.9 S_e^2$, $k_{rn}=0.5 (1- S_e)$}\\
			{Injection rate} & \multicolumn{2}{c}{0.05 PV/year}\\
			\multirow{6}*{Boundary conditions} & \multicolumn{2}{c}{$\phi_w(0,y,z,t)=\phi_n(0,y,z,t)=0$ kg/(m$^2$s)}\\
			& \multicolumn{2}{c}{$\phi_w(200,y,t)=\phi_n(200,y,z,t)=0$ kg/(m$^2$s)}\\
			& \multicolumn{2}{c}{$\phi_w(x,0,z,t)=\phi_n(x,0,z,t)=0$ kg/(m$^2$s)}\\
			& \multicolumn{2}{c}{$\phi_w(x,200,z,t)=\phi_n(x,200,z,t)=0$ kg/(m$^2$s)}\\
			& \multicolumn{2}{c}{$S_w(x,y,0,t) = 1-S_{nr}$, $p_n(x,y,0,t)=5\times10^{7}$ Pa}\\
			& \multicolumn{2}{c}{$\phi_w(x,y,200,t)=\phi_n(x,y,200,t)=0$ kg/(m$^2$s)}\\
			{Initial conditions} & \multicolumn{2}{c}{$S_w(x,y,z,0)=1-S_{nr}$}\\
			\bottomrule
		\end{tabular}
		\label{case5}
	\end{center}
\end{table}
The result of the strong scalability test for Case-5 is shown in Table \ref{case5_strong}. In this test, the mesh size is $256\times256\times256$, resulting in 33,949,186 DoF. With the increase in number of processors, the total execution time($T_{\text{total}}$), execution time per LIMEX step($t_{\text{LIMEX}}$), and execution time per linear iteration($t_{\text{linear}}$) scaling well. 

\begin{table}[H]
	\begin{center}
		\caption{Strong scaling test for Case-5}
		\begin{tabular}{*{7}{c}}
			\toprule
			{PE} & {$T_{\text{total}}$(s)} & {$N_{\text{LIMEX}}$} & {$t_{\text{LIMEX}}$(s)} & {$N_{\text{linear}}$} &  {$t_{\text{linear}}$(s)} & {Speedup}\\
			\midrule
			{256} & {1182.3} & {67+2} & {17.13} & {458} & {0.335} & {1}\\
			{512} & {640.0} & {67+2} & {9.28} & {489} & {0.194} & {1.847}\\
			{1024} & {372.2} & {67+2} & {5.39} & {478} & {0.146} & {1.719}\\
			\bottomrule
		\end{tabular}
		\label{case5_strong}
	\end{center}
\end{table}

The result of the weak scalability test for Case-5 is shown in Table\ref{case5_weak}. The mesh size increases from $256\times256\times256$ to $1024\times1024\times1024$ with the increase of the number of processors from 64 to 4096. For this highly nonlinear multiphase flow problem, LIMEX takes more adaptive time steps to capture more phenomena with mesh refining. The execution time increased from 3606.7s to 5587.0s only, when the DoF increased 63 times.
\begin{table}[H]
	\begin{center}
		\caption{Weak scaling test for Case-5}
		\begin{tabular}{*{7}{c}}
			\toprule
			{PE} & {DoF} & {$T_{\text{total}}$(s)} & {$N_{\text{LIMEX}}$} & {$t_{\text{LIMEX}}$(s)} & {$N_{\text{linear}}$} & {$t_{\text{linear}}$(s)}\\
			\midrule
			{64} & {33,949,186} & {3606.7} & {67+2} & {52.3} & {458} & {0.559}\\
			{512} & {270,011,394} & {4248.6} & {72+2} & {57.4} & {576} & {0.746}\\
			{4096} & {2,153,781,250} & {5587.0} & {79+2} & {69.0} & {599} & {1.324}\\
			\bottomrule
		\end{tabular}
		\label{case5_weak}
	\end{center}
\end{table}

\section{Conclusion}
\label{}
In this paper, a highly scalable framework for simulations of multi-phase flow in heterogeneous porous media with capillary and gravity is proposed. The applied fully coupled and fully implicit scheme gets rid of the restriction on the time step size and maximizes the stability of the proposed framework. The adaptive time stepping scheme is achieved by adapting LIMEX with the error estimator. The accuracy and efficiency of the proposed framework are validated by several numerical cases. The great scalability of the proposed framework is verified by the scaling tests up to 4096 processors with 2 billion DoFs.

\section*{Acknowledgements}
The author would like to express their deep gratitude to Dr. Dmitry Logashenko, Dr. Arne N{\"a}gel, Prof. Stephan Matthai, Prof. Gabriel Wittum, Prof. Dr. Hussein Hoteit, and Prof. Dr. S. Majid Hassanizadeh for fruitful discussions and essential suggestions. For the numerical tests, the authors used the Shaheen III supercomputer managed by the Supercomputing Core Laboratory at KAUST in Thuwal, Saudi Arabia. The author thanks the KAUST HPC support team for their assistance with this equipment.

\bibliographystyle{unsrt}  
\bibliography{references}  


\end{document}